\documentclass[12pt]{article}
\usepackage{rotating}
\usepackage{multirow}
\usepackage{times}
\usepackage{cite}
\usepackage{amsmath}
\usepackage{fullpage}
\usepackage{setspace}
\usepackage{booktabs}
\usepackage{graphicx}
\usepackage{color}
\usepackage{float}

\usepackage[top=0.85in,left=1in,right=1in,footskip=0.75in]{geometry}

\begin{document}
\vspace*{0.2in}

\begin{flushleft}
{\Large
\textbf\newline{Untangling the hairball: fitness based asymptotic reduction of biological networks} 
}
\newline
\\
F Proulx-Giraldeau   \textsuperscript{*},
TJ Rademaker \textsuperscript{*},
P Fran\c cois
\\
\bigskip

%
%
* Equal Contribution 


\end{flushleft}
\section*{Abstract}

Complex mathematical models of interaction networks are routinely used for prediction in systems biology. However, it is difficult to reconcile network complexities with a formal understanding of their behavior. Here, we propose a simple procedure  (called $\bar \phi$) to reduce biological models to functional submodules, using statistical mechanics of complex systems combined with a fitness-based approach inspired by {\it in silico} evolution. $\bar \phi$ works by putting parameters or combination of parameters to some asymptotic limit, while keeping (or slightly improving) the model performance, and requires parameter symmetry breaking for more complex models. We illustrate $\bar \phi$ on biochemical adaptation and on different models of immune recognition by T cells. An intractable model of immune recognition with close to a hundred individual transition rates is reduced to a simple two-parameter model. $\bar \phi$ extracts three different mechanisms for early immune recognition, and automatically discovers similar functional modules in different models of the same process, allowing for model classification and comparison. Our procedure can be applied to biological networks based on rate equations using a fitness function that quantifies phenotypic performance.



\section*{Introduction}

As more and more systems-level data are becoming available, new modelling approaches have been developed to tackle biological complexity. A popular bottom-up route inspired by ``-omics" aims at exhaustively describing and modelling parameters and interactions  \cite{Karr:2012bh,Markram:2015dq}. The underlying assumption is that the behavior of systems taken as a whole will naturally emerge from the modelling of its underlying parts. While such approaches are rooted in biological realism, there are  well-known modelling issues. By design, complex models are challenging to study and to use. More fundamentally,  connectomics does not necessarily yield clear functional information of the ensemble, as recently exemplified in neuroscience \cite{Jonas:2017fw}.  Big models are also prone to overfitting \cite{Mayer:2010kb,Lever:2016hm}, which undermines their predictive power.  It is thus not clear how to tackle network complexity in a predictive way, or, to quote Gunawardena \cite{Gunawardena:2013}   `` how the biological wood emerges from the molecular trees".

More synthetic approaches have actually proved successful. Biological networks are known to be modular \cite{Milo:2002}, suggesting that much of the biological complexity  emerges from the combinatorics of simple functional modules.  Specific examples from immunology to embryonic development have shown that small and well-designed phenotypic networks can  recapitulate most important properties of complex networks \cite{Corson:2012, Francois:2013, Lever:2014}. A fundamental argument in favor of such ``phenotypic modelling'' is that biochemical networks themselves are not necessarily conserved, while their function is. This is exemplified by the significant network differences in segmentation of different vertebrates despite very similar functional roles and dynamics \cite{Krol:2011}. It suggests that the level of the phenotype is the most appropriate one and that a too detailed (gene-centric) view might not be the best level to assess systems as a whole.
  
 The predictive power of simple models has been theoretically studied by Sethna and co-workers, who argued that even without complete knowledge of parameters, one is able to  fit experimental data and predict new behavior \cite{Brown:2003ta,Brown:2004kt, Gutenkunst:2007,Transtrum:2015ts}. These ideas are inspired by recent progress in statistical physics, where ``parameter space compression'' naturally occurs, so that dynamics of complex systems can actually be well described with few effective parameters \cite{Machta:2013ga}. Methods have further been developed to generate parsimonious models based on data fitting that are able to make new predictions \cite{Daniels:2015gi,Daniels:2015eg}. However such simplified models might not be easily connected to actual biological networks. An alternative strategy is to enumerate \cite{Ma:2009,Cotterell:2010hk} or evolve {\it in silico}  networks that perform complex biological functions \cite{Francois:2014}, using predefined biochemical grammar, and allowing for a more direct comparison with actual biology. Such approaches typically give many results. However common network features can be identified in retrospect and as such are predictive of biology \cite{Francois:2014}. Nevertheless, as soon as a microscopic network-based formalism is chosen, tedious labor is required to identify and study underlying principles and dynamics.  If we had a systematic method to simplify/coarse-grain models of networks while preserving their functions, we could better understand, compare and classify different models. This would  allow us to extract dynamic principles underlying given phenotypes with maximum predictive power .

Inspired by a recently proposed boundary manifold approach   \cite{Transtrum:2014wd}, we propose a simple method to coarse-grain phenotypic models, focusing on their functional properties via the definition of a so-called fitness. Complex networks, described by rate equations, are then reduced to much simpler ones that perform the same biological function. We first reduce  biochemical adaptation, then consider the more  challenging problem of absolute discrimination, an important instance being the early immune recognition \cite{Francois:2016ig}. In particular, we succeed in identifying functional and mathematical correspondence between different models of the same process. By categorizing and classifying them, we identify general principles and biological constraints for absolute discrimination. Our approach suggests that complex models can indeed be studied and compared using parameter reduction, and that minimal phenotypic models can be systematically generated from more complex ones. This may significantly enhance our understanding of biological dynamics from a complex network description.

\section*{Materials and methods}
\subsection*{An algorithm for fitness based asymptotic reduction}

Transtrum \& Qiu  \cite{Transtrum:2014wd, Transtrum:2016jm} studied the problem of data fitting using cellular regulatory networks modelled as coupled ordinary differential equations. They proposed that models can be reduced by following geodesics in parameter space, using error fitting as the basis for the metric. This defines the Manifold Boundary Approximation Method (abbreviated as MBAM) that extracts the minimum number of parameters compatible with data  \cite{Transtrum:2014wd}. 

While simplifying models to fit data is crucial, it would also be useful to have a more synthetic approach to isolate and identify functional parts of networks. This would be especially useful for model comparison of processes where abstract functional features of the models (e.g. the qualitative shape of a response) might not correspond to one another, or where the underlying networks are different while they perform the same overall function \cite{Krol:2011}. 
We thus elaborate on the  approach of  \cite{Transtrum:2014wd} and describe in the following an algorithm for FItness Based Asymptotic parameter Reduction (abbreviated as FIBAR or $\bar \phi$). $\bar \phi$  does not aim at fitting data, but focuses on extracting functional networks, associated to a given biological function. To define biological function, we  require a general fitness (symbolized by $\phi$) to quantify performance. Fitness is broadly defined as a mathematical quantity encoding biological function in an almost parameter independent way, which allows for a much broader search in parameter space than traditional data fitting (examples are given in the next sections).   The term fitness is   inspired by its use in  evolutionary algorithms to select for coarse-grained functional networks \cite{Francois:2014}. We then define model reduction as the search for networks with as few parameters as possible optimizing  a predefined fitness. There is no reason \textit{a priori} that such a procedure would converge for arbitrary networks or fitness functions: it might simply not be possible to optimize a fitness without some preexisting network features. A more traditional route to optimization would rather be to increase the number of parameters to explore missing dimensions, rather than decrease them (see discussions in  \cite{Daniels:2015gi,Daniels:2015eg}) . We will show how $\bar \phi$ reveals network features in known models that were explicitly designed to perform the fitness of interest.

Due to the absence of an explicit cost function to fit data, there is no equivalence in $\bar \phi$ to the metric in parameter space in the MBAM allowing to incrementally update parameters. However, upon further inspection, it appears that most limits in  \cite{Transtrum:2014wd} correspond to simple transformations in parameter space:  single parameters disappear by putting them to $0$ or $\infty$, or by taking limits where their product or ratio are constant while individual parameters  go to  $0$ or $\infty$. In retrospect, some of these transformations can  be interpreted as well-known limits such as quasi-static assumptions or dimensionless reduction, but there are more subtle transformations, as will appear below.  

Instead of computing geodesics in parameter space, we directly probe asymptotic limits for all parameters, either singly or in pair. Practically, we generate a new parameter set by multiplying and dividing a parameter by a large enough rescaling factor $f$ (which is a parameter of our algorithm, we have taken $f=10$ for the simulations presented here), keeping all other parameters constant, or doing the same operation on a couple of parameters.

At each step of the algorithm, we compute the behavior of the network when changing single parameters, or any couple of parameters by factor $f$ in both directions. We then compute the change of fitness for each of the new models with changed parameters. In most cases, there are parameter modifications that leave the fitness unchanged or even slightly improve network behavior. Among this ensemble, we follow a conservative approach and select (randomly or deterministically) one set of parameter modifications that minimizes the fitness change. We then implement parameter reduction by effectively pushing the corresponding parameters to $0$ or $\infty$, and iterate the method until no further reduction enhances the fitness or leaves it unchanged, or until all parameters are reduced. The evaluation of these limits effectively removes parameters from the system while keeping  the fitness unchanged or incrementally improving it. There are technical issues we have to consider: for instance, if two parameters go to $\infty$ some numerical choices have to be made about the best way to implement this. Our choice was to keep the reduction simple : in this example, instead of defining explicitly a new parameter, we increase both parameters to a very high value, freeze one of them, and  allow variation of the other one for subsequent steps of the algorithm. Another issue with asymptotic limits for rates is that corresponding divergence of variables might occur.  To ensure proper network behavior, we thus impose overall mass conservation for some predefined variables, e.g. total concentration of an enzyme (which effectively adds fluxes to the free form of the considered biochemical species). We also explicitly test for convergence of differential equations and discard parameter modifications leading to numerical divergences. Details on the implementation of the reduction rules for specific models are presented in the Supplement and can be automatically implemented for any model based on rate equations.

\begin{figure}
\includegraphics[width=6.5 in]{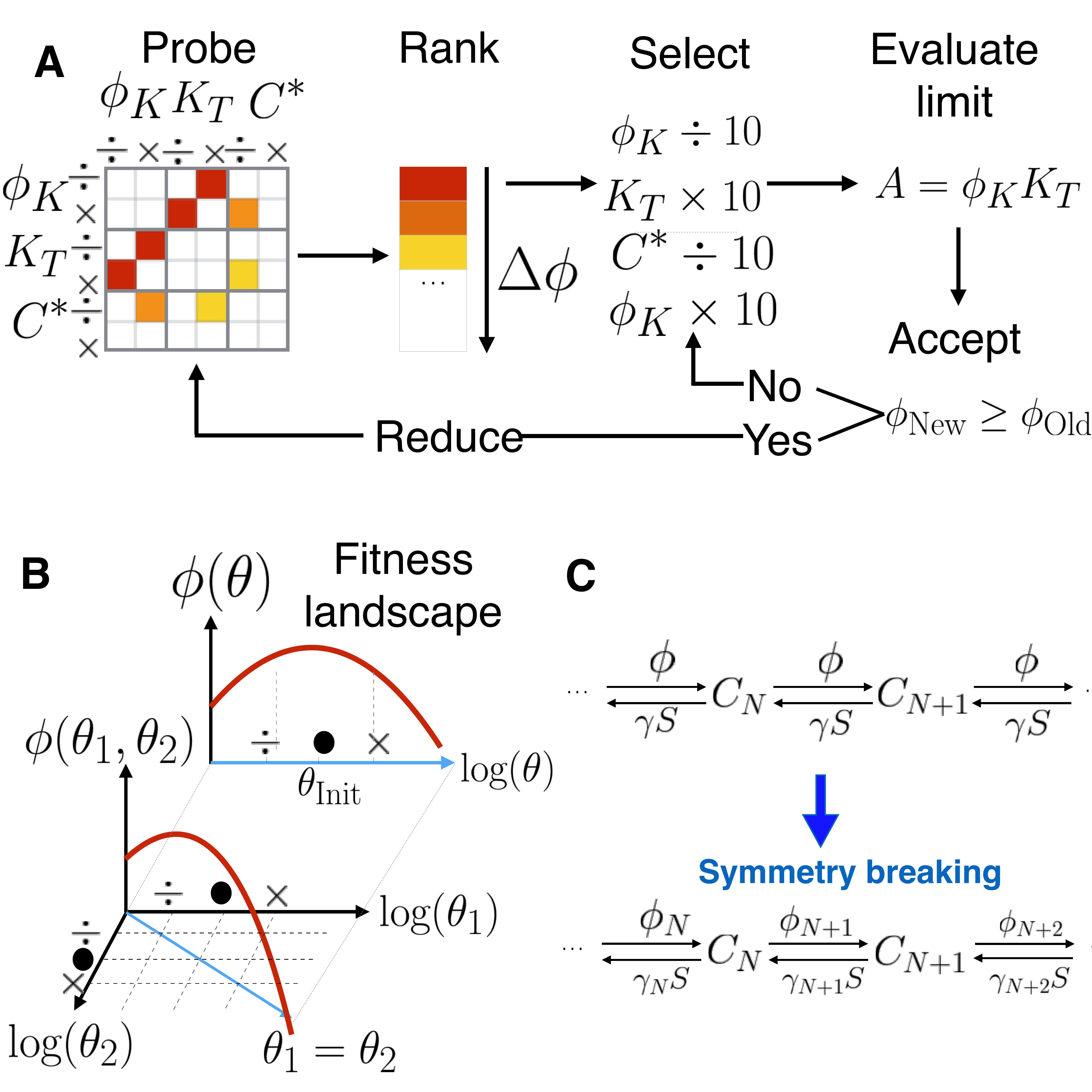}
\caption{ Summary of $\bar \phi$ algorithm. (A) Asymptotic fitness evaluation and reduction: for a given network, the values of fitness $\phi$ are computed for asymptotic values of  parameters or couples of parameters. If the fitness is improved (warmer colors), one subset of improving parameters is chosen and pushed to its corresponding limits, effectively reducing the number of parameters. This process is iterated.  See main text for details. (B) Parameter symmetry breaking:  a given parameter present in multiple rate equations (here $\theta$) is turned into multiple parameters ($\theta_1, \theta_2$) that can be varied independently during asymptotic fitness evaluation. (C) Examples of parameter symmetry breaking, considering a biochemical cascade similar to the model from \cite{Francois:2013}. See main text for comments.} \label{fig-algo}
\end{figure}

These iterations of parameter changes alone do not always lead to simpler networks. This is also observed in the MBAM when it is sometimes no longer possible to fit all data as well upon parameter reduction. However, with the goal to extract minimal functional networks, we can circumvent this problem by implementing  what we call ``symmetry breaking" of the parameters (Fig. \ref{fig-algo} B-C): in most networks, different biochemical reactions are assumed to be controlled by the same parameter. An example is a kinase acting on different complexes in a proofreading cascade with the same reaction rate. However, an alternative hypothesis is that certain steps in the cascade are recognized to activate specific pathways, or targeted for removal (e.g. in ``limited signalling models", the signalling step is specifically tagged, thus having dual specificity \cite{Lever:2014}). So to further reduce parameters, we assume that those rates, which are initially equal, can now be varied independently by $\bar \phi$ (Fig. \ref{fig-algo} C). Symmetry breaking in parameter space allow us to reduce models to a few relevant parameters/equations, and as explained below are necessary to extract simple descriptions of network functions. Note that symmetry breaking transiently expand the number of parameters, allowing for a more global search for a reduced model in the complex space of networks. Fig. \ref{fig-algo} A summarizes this asymptotic reduction.

We have implemented $\bar \phi$ in MATLAB for the specific cases described here, and samples of code used are available as Supplementary Materials.

\subsection*{Defining the fitness}

To illustrate the $\bar \phi$ algorithm, we apply it to two different biological problems:  biochemical adaptation and absolute discrimination.  In this section we briefly describe those problems and define the associated fitness functions.

\begin{figure}
\includegraphics[width=6.5 in]{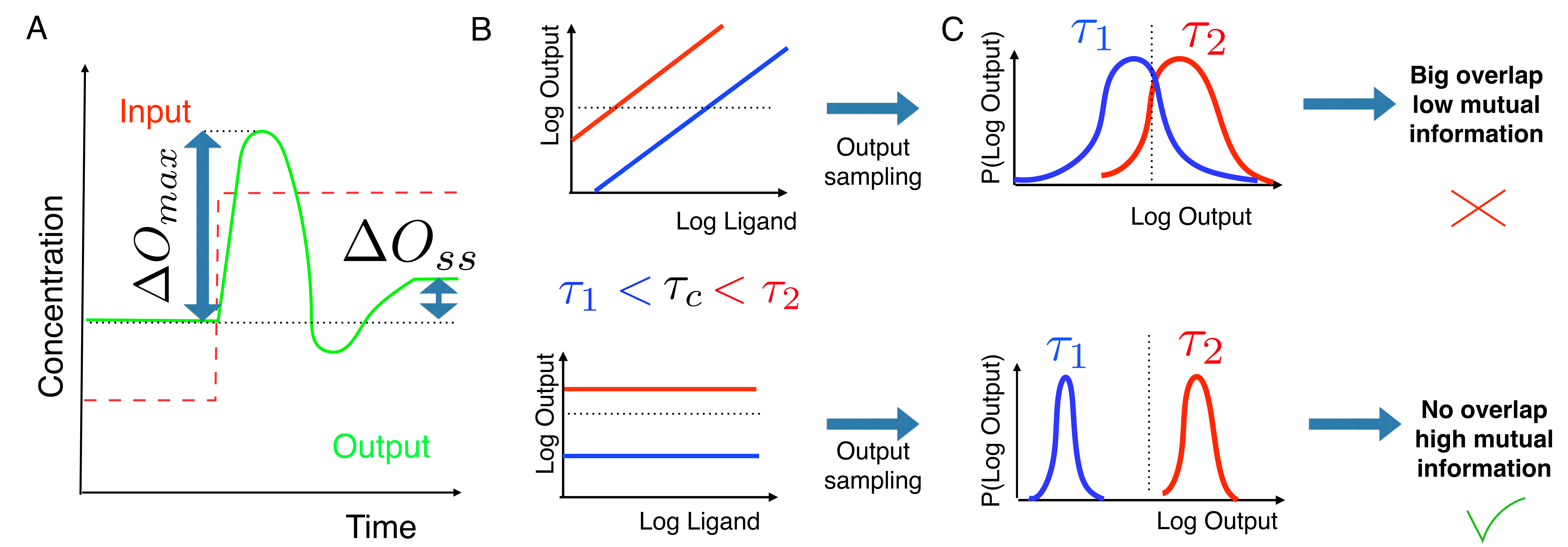}
\caption{ Fitness explanations. (A) Fitness used for biochemical adaptation. Step of an Input variable is imposed (red dashed line) and behavior of an Output variable is computed (green line). Maximum deviation $\Delta O_{max}$ and steady state deviation $\Delta O_{ss}$ are measured and optimized for fitness computation. (B) Schematics of response line for absolute discrimination. We represent expected dose response curves for a ``bad'' (top) and a ``good'' (bottom) model . Response to different binding times $\tau$ are symbolized by different colors.  For the ``bad'' monotonic model (e.g. kinetic proofreading \cite{McKeithan:1995}),  by setting a threshold (horizontal dashed line), multiple intersections with different lines corresponding to different $\tau$s are found, which means it is not possible to measure $\tau$  based on the Output. Bottom corresponds to absolute discrimination: flat responses  plateau at different Output values easily measure $\tau$. Thus, the network can easily decide the position of $\tau$ with respect to a given threshold  (horizontal dashed line). (C) For actual fitness computation, we sample the possible values of the Output with respect to a predefined Ligand distribution for different $\tau$s (we have indicated threshold similar to panel (B) by a dahsed line). If the distribution are not well separated, one can not discriminate between $\tau$s  based on Outputs and mutual information between Output and $\tau$ is low. If they are well separated, one can discriminate $\tau$s based on Output and mutual information is high. See technical details in the Supplement.} \label{fig-fitness}
\end{figure}

The first problem we study is biochemical adaptation, a classical, ubiquitous phenomenon in biology in which an output variable returns to a fixed homeostatic value after a change of Input (see Fig. \ref{fig-fitness} A).  We apply $\bar \phi$ on models inspired by \cite{Ma:2009, Transtrum:2016jm},  expanding Michaelis-Menten approximations into additional rate equations, which further allows  to account for some implicit constraints of the original models, see details in the Supplement.  We use a fitness that is first detailed in \cite{Francois:2008}: we measure the deviations from equilibrium at steady state $\Delta O_{ss}$ and the maximum deviation $\Delta O_{max}$ after a change of Input, and aim at minimizing the former while maximizing the latter.  Combining both numbers into a single sum   $\Delta O_{max}+\epsilon/\Delta O_{ss}$  gives the fitness we are maximizing (see more details in the Supplement). This simple case study illustrates how $\bar \phi$ works and allows us to compare our findings to previous work such as \cite{Transtrum:2016jm}.

 The second problem is absolute discrimination, defined as the sensitive and specific recognition of signalling ligands based on one biochemical parameter. Possible instances of this problem can be found in immune recognition between self and not self  for T cells \cite{Feinerman:2008b,Francois:2016ig} or mast cells  \cite{Torigoe:1998vj}, and recent works using chimeric DNA receptor confirm sharp thresholding based on binding times \cite{Taylor:2016eo}. More precisely, we consider models where a cell is exposed  to an amount $L$ of identical ligands, where their binding time $\tau$ defines their quality. Then the cell should discriminate only on $\tau$, i.e. it should decide if $\tau$ is higher or lower than a critical value $\tau_c$ {\it independently} of ligand concentration $L$. This is a nontrivial problem, since many ligands with binding time slightly lower than $\tau_c$ should not trigger a response, while few ligands with binding time slightly higher than $\tau_c$ should.  Absolute discrimination has direct biomedical relevance, which explains why there are  models of various complexities,  encompassing several interesting and generic features of biochemical network (biochemical adaptation, proofreading, positive and negative feedback loops, combinatorics, etc.). Such  models serve as ideal tests for the generality of $\bar \phi$.

The performance of a network performing absolute discrimination is illustrated in  Fig. \ref{fig-fitness}. We can plot the values of the network output $O$ as a function of ligand concentration $L$, for different values of $\tau$ (Fig. \ref{fig-fitness} B). Absolute discrimination between ligands is possible only if one (or more realistically few) values of $\tau$ correspond to a given Output value $O(L,\tau)$ (as detailed in \cite{Francois:2016ig}). Intuitively, this is not possible if the dose response curves $O(L,\tau)$ are monotonic: the reason is that for any value of output $O$, one can find many associated couples of $(L,\tau)$ (see Fig. \ref{fig-fitness} B). Thus, ideal performance corresponds to separated horizontal lines, encoding different values of $O$ for different $\tau$ independently of $L$ (Fig. \ref{fig-fitness} B). For suboptimal cases and optimization purposes, a probabilistic framework is useful. Our fitness is the mutual information between the distribution of outputs $O$ with $\tau$ for a predefined sampling of $L$, as proposed in \cite{Lalanne:2013}. If those distributions are not well separated (meaning that we can frequently observe the same Output value for different values of $\tau$ and $L$,  Fig. \ref{fig-fitness} C top), the mutual information is low and the network performance is bad. Conversely, if those distributions are well separated ( Fig. \ref{fig-fitness} C bottom), this means that a given Output value is statistically very often associated to a given value of $\tau$. Then the mutual information is high and network performance is good. More details on this computation can be found in  the Supplement (Fig. S2).

We have run $\bar\phi$ on three different models of this process: ``adaptive sorting" with one proofreading step \cite{Lalanne:2013}, a simple model based on feedback by phosphatase SHP-1 from \cite{Francois:2013} (``SHP-1 model"), and a complex realistic  model accounting for multiple feedbacks from \cite{Lipniacki:2008} (``Lipniacki model").  Initial models are described in more details in following sections. We have taken published parameters as initial conditions.  Those three models were all explicitly designed to describe absolute discrimination, modelled as sensitive and specific sensing of ligands of a given binding time $\tau$ \cite{Francois:2016ig}, so ideally those networks would have perfect fitness. However due to various biochemical constraints, these three models have very good initial  (but not necessarily perfect) performance  for absolute discrimination. We see that after some initial fitness improvement,  $\bar \phi$  reaches an optimum fitness within a few steps and thus merely simplifies models while keeping constant fitness (see fitness values in the Supplement). We have tested $\bar \phi$ with several parameters of the fitness functions, and we give in the following for each model the most simplified networks obtained with the help of those fitness functions. Complementary details and other reductions are given in the Supplement.
  
For both problems, $\bar \phi$ succeeds in fully reducing the system to a single equation with essentially two effective parameters (see Tables in the Supplement, final model is given in the FINAL OUTPUT formula, and discussion of the effective parameters in the section ``Comparison and categorization of models''). However, to help understanding the mathematical structure of the models,  it is helpful to deconvolve some of the reduction steps from the final model. In particular, this helps to identify functional submodules of the network that perform independent computations. Thus for each example  below, we give a small set of differential equations capturing the functional mechanisms of the reduced model . In Figures we show in the ``FINAL'' panel the behaviour of the full system of ODEs including all parameters (but potentially very big or very small values after reduction), and thus including local flux conservation.

\section*{Results}

\subsection*{$\bar \phi$ for biochemical adaptation: feedforward and feedback models}

\begin{figure}
\includegraphics[width=6.5 in]{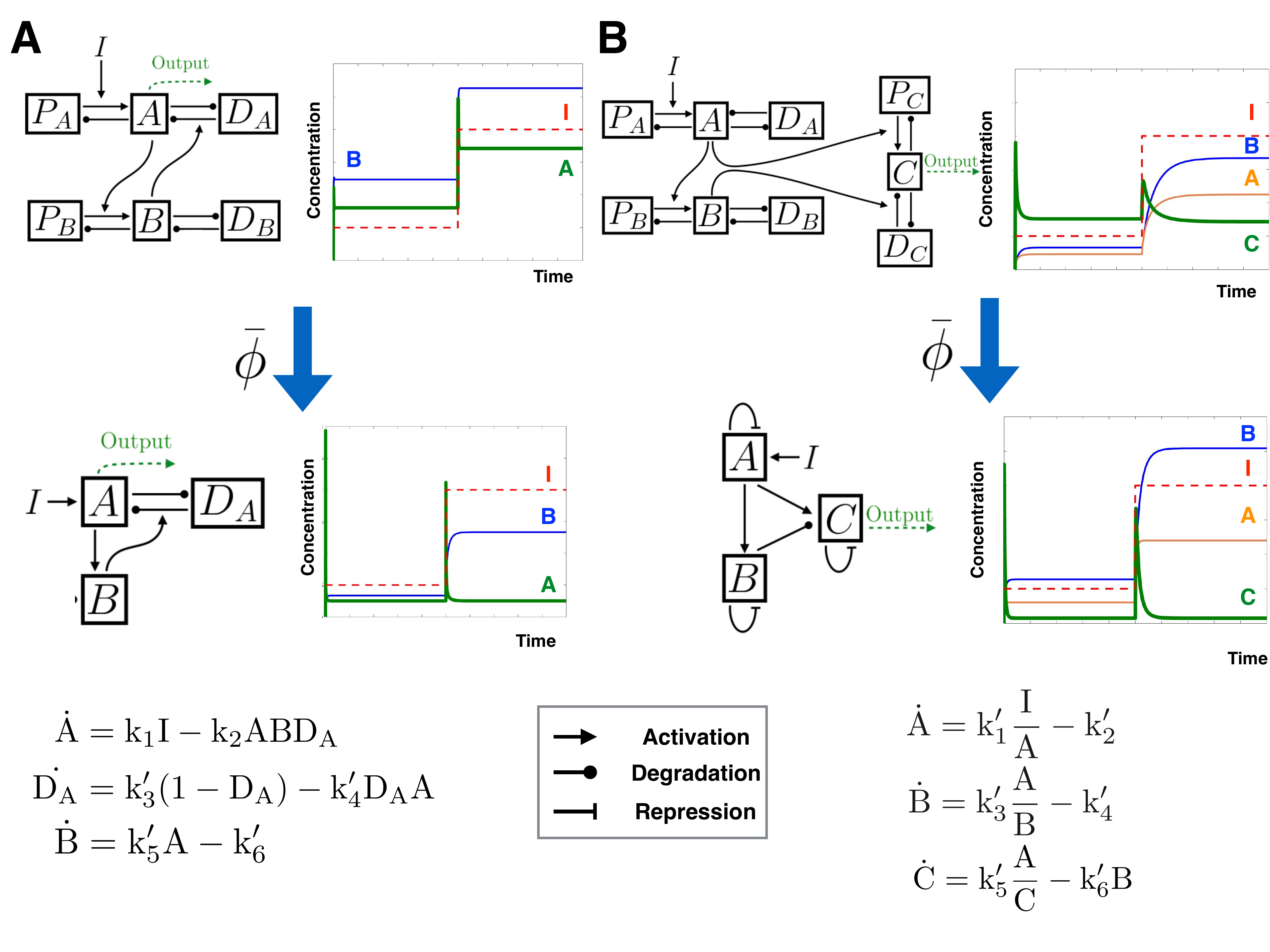}
\caption{Adaptation networks considered and their reduction by $\bar \phi$. We explicitly include Production and Degradation nodes ($P$s and $D$s) that are directly reduced into Michaelis-Menten kinetics in other works. From top to bottom, we show the original network, the reduced network, and the equations for the reduced network. Dynamics of the networks under control of a step input ($I$) is also shown.  Notice that the initial networks are not adaptive while the final reduced networks are. (A) Negative feedback network, including enzymes responsible for Michaelis-Menten kinetics for production and degradation. $A$ is the adaptive variable. (B) Incoherent feedforward networks.  $C$ is the adaptive variable. }\label{fig-adapt}
\end{figure}

The problem of biochemical adaptation allows us to simply illustrate and compare the algorithm on problems described and well-studied elsewhere. We consider two models based on feedforward and feedback loops, with corresponding interactions between the nodes. These models are adapted from  \cite{Ma:2009}, and  have network topologies known to be compatible with biochemical adaptation.  $\bar \phi$ is designed to work with rate equations, so to keep mathematical expressions compatible with the ones in \cite{Ma:2009} we have to introduce new nodes corresponding to enzymes and regulations for production and degradation. For instance, a nonlinear degradation rate for protein $A$ of the form $A/(A+A_0)$ in  \cite{Ma:2009} implicitly means that $A$ deactivates its own degrading enzyme, that we include and call $D_A$ ( see equations in the Supplement). This gives networks with 6 differential equations/ 12 parameters for the negative feedback network, and 9 differential equations/18 parameters for incoherent feedforward network.  For this problem, we have not tried to optimize initial parameters for the networks, instead we start with arbitrary parameters (and thus arbitrary non-adaptive behaviour) and we simply let $\bar \phi$ reduce the system using the fitness function defined above. The goal is to test efficiency of  $\bar \phi$,  and to see if it finds a subset of nodes/parameters compatible with biochemical adaptation by pure parameter reduction (we know from analytical studies similar to what is done in  \cite{Ma:2009} that such solutions exist but it is not clear that they can be found directly by asymptotic parameter reduction).  Fig. \ref{fig-adapt} summarizes the initial network topologies considered, including the associated enzymes and the final reduced models, with equations. Steps of the reductions are given in the Supplement. 

Both networks  converge towards adaptation working in a very similar way to networks previously described in  \cite{Ma:2009, Transtrum:2016jm}. For the negative feedback network of Fig. \ref{fig-adapt} A, at steady state, $A$ is pinned to a value independent of $I$ ensuring its adaptation by stationarity of protein $B$ ($\dot B=0$). Stationarity of $A$ imposes that $B$ essentially buffers the Input variation and that $A$ transiently feels $I$ (see equations and corresponding behavior on  Fig. \ref{fig-adapt} A). This is a classical implementation of integral feedback \cite{Yi:2000tj} with a minimum number of two nodes, automatically rediscovered by $\bar \phi$.

We see similar behavior for reduction of the incoherent feedforward networks, Fig. \ref{fig-adapt} B. At steady state, stationarity of $B$ pins the ratio $A/B$ to a value independent of $I$, while stationarity of $C$ imposes that $C$ is proportional to $A/B$ and thus adaptive  (see equations and corresponding behavior in Fig. \ref{fig-adapt} B). This is a classical implementation of another feedforward adaptive system  \cite{Francois:2008,Ma:2009}, rediscovered by $\bar \phi$.  When varying simulation parameters for $\bar \phi$, we can see some variability in the results, where steady state relations between $A,B,C$ are formally identical but with another logic (see details of such a reduction in the Supplement).

During parameter reduction, ratios of parameters are systematically eliminated, corresponding to classical limits such as saturation or dimensionless reductions, as detailed in the Supplement. Similar limits were observed in \cite{Transtrum:2016jm} when applying the MBAM to fits simulated data for biochemical adaptation. The systems reduce in both cases to a minimum number of differential equations, allowing for transient dynamics of the adaptive variable. Interestingly, in this case we have not attempted to optimize parameters \textit{a priori}, but nevertheless $\bar \phi$ is able to converge towards adaptive behaviour only by removing parameters. In the end, we recover known reduced models for biochemical adaptation, very similar to what is obtained with artificial data fitting in \cite{Transtrum:2016jm}, confirming the efficiency and robustness of fitness based asymptotic reduction.

\begin{figure}
\includegraphics[width=6.5 in]{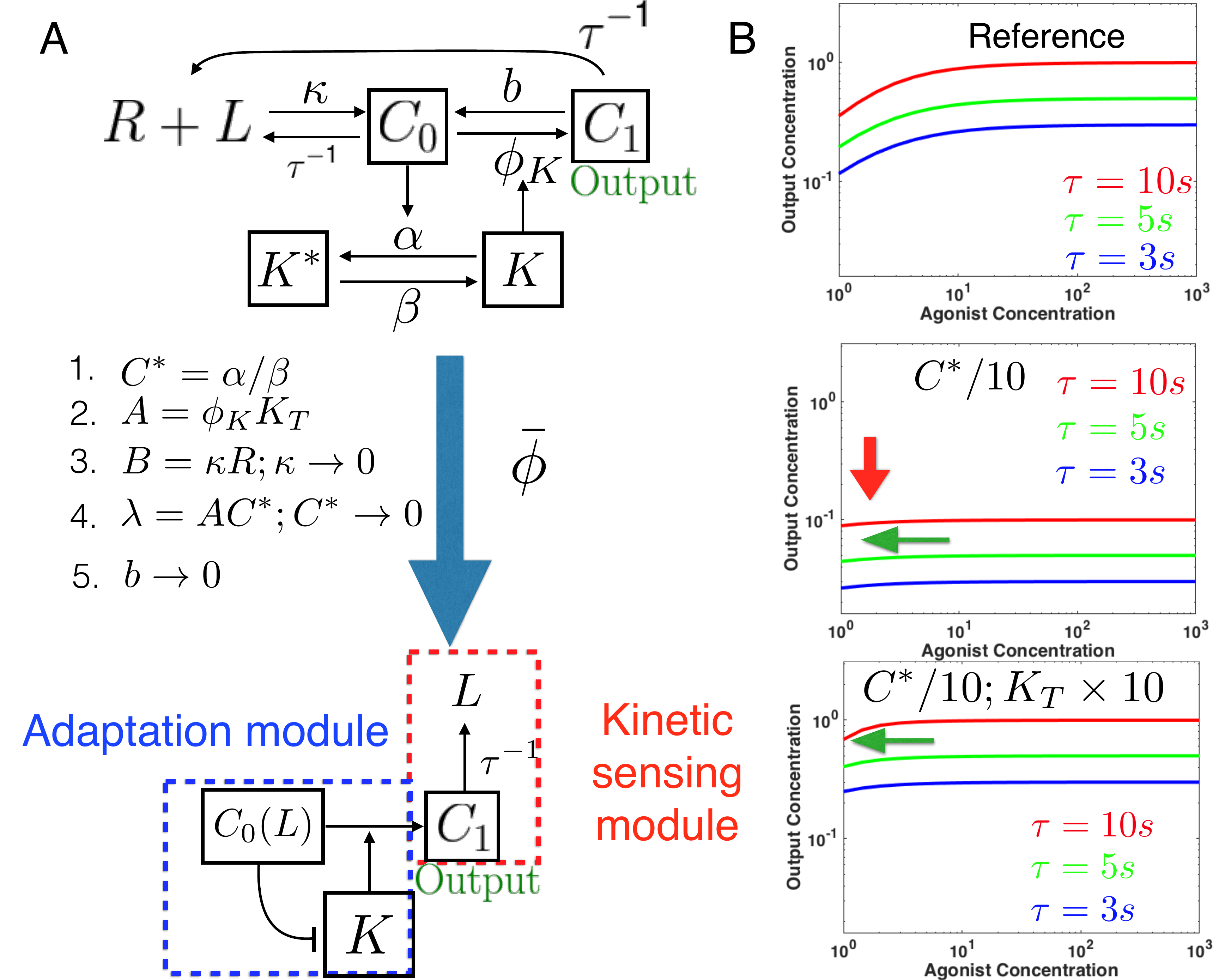}
\caption{ Reduction of Adaptive sorting. (A) Sketch of the network, with 5 steps of reductions by $\bar\phi$. Adaptation and kinetic sensing modules are indicated for comparison with reduction of other models. (B) Illustration of the specificity/response trade-off solved by Step 4 of $\bar \phi$. Compared to the reference behavior (top panel), decreasing $C^*$ (middle panel) increases specificity with less $L$ dependency (horizontal green arrow) but globally reduces signal (vertical red arrow). If $K_T$ is simultaneously increased (bottom panel), specificity alone is increased without detrimental effect on overall response, which is the path found  by $\bar \phi$.}\label{fig-AS}
\end{figure}

\subsubsection*{$\bar \phi$ for  adaptive sorting}

We now proceed with applications of  $\bar \phi$ to  the more challenging problem of absolute discrimination. Adaptive sorting \cite{Lalanne:2013} is one of the simplest models of absolute discrimination. It consists of a one-step kinetic proofreading cascade  \cite{McKeithan:1995} (converting complex $C_0$ into $C_1$) combined to a negative feedforward interaction mediated by a kinase $K$, see Fig. \ref{fig-AS} A for an illustration.  A biological realization of adaptive sorting exists for FCR receptors \cite{Torigoe:1998vj}. 

This model has a complete analytic description in the limit where the backward rate from $C_1$ to $C_0$ cancels out \cite{Lalanne:2013}. The dynamics of $C_1$ is then given by:

\begin{equation}
\dot C_1=\phi_K K C_0(L) -\tau^{-1}C_1 \qquad \textrm{with} \qquad K= K_T \frac{C^*}{C_0(L)+C^*}   \label{C1}
\end{equation}

 $K$ is the activity of a kinase regulated by complex $C_0(L)$, itself proportional to ligand concentration $L$. $K$ activity is repressed by $C_0$ (Fig. \ref{fig-AS}, Eq. \ref{C1}), implementing an incoherent feedforward loop in the network (full system of equations are given in the Supplement).
 
 Absolute discrimination is possible when $C_1$  is a pure function of $\tau$ irrespective of $L$ (so that $C_1$ encodes $\tau$ directly) as discussed in \cite{Lalanne:2013, Francois:2016ig}. \textit{A priori}, both $C_0$ and $C_1$ depend on the input ligand concentration $L$.   If we require $C_1$ to be independent of $L$, the product $KC_0$ has to become a constant irrespective of $L$. This is possible because $K$ is repressed by $C_0$, so there is a ``tug-of-war"  on $C_1$ production between the substrate concentration $C_0$, and its negative effect on $K$. In the limit of large enough $C_0$, $K$ is indeed becoming inversely proportional to $C_0$, giving a production rate of $C_1$ independent of $L$. $\tau$ dependency is then encoded in the dissociation rate of $C_1$ so that in the end $C_1$ is a pure function of $\tau$.

The steps of  $\bar \phi$  for adaptive sorting are summarized in Fig. \ref{fig-AS} A.  The first steps correspond to standard operations: step 1 is a quasi-static assumption on kinase concentration, step 2 brings together parameters having similar influence on the behavior,  and step 3 is  equivalent to assuming receptors are never saturated. Those steps are already taken in \cite{Lalanne:2013}, and are automatically rediscovered by $\bar \phi$.  Notably, we see that during reduction several effective parameters emerge, e.g. parameter $A=K_T \phi$ can be identified in retrospect as the maximum possible activity of kinase K.

Step 4 is the most interesting step and corresponds to a nontrivial parameter modification specific to $\bar \phi$, which simultaneously reinforces the two tug-of-war terms described above, so that they balance more efficiently. This transformation solves a trade-off between sensitivity of the network and magnitude in response, illustrated in Fig. \ref{fig-AS} B. If one decreases only parameter $C^*$, the dose response curves for different $\tau$s become flatter, allowing for better separation of $\tau$s (i.e. specificity),  Fig. \ref{fig-AS} B, middle panel.  However, the magnitude of the dose response curves is proportional to $C^*$ so that if we were to take $C^*=0$, all dose response curves would go to $0$ as well and the network would lose its ability to respond. It is only when both $C^*$ and the parameter $A=K_T \phi_K$ are changed in concert that we can increase specificity without losing response, Fig. \ref{fig-AS} B, bottom panel. This ensures that $K(L)$ becomes always proportional to $L$ without changing the maximum production rate  $A C^*$ of $C_1$. $\bar \phi$ finalizes the reduction by putting other parameters to  limits that do not significantly change $C_1$'s value. There is no need to perform symmetry breaking for this model to reach optimal behavior and one-parameter reduction.

This simple example illustrates that  not only is $\bar \phi$ able to rediscover automatically classical reduction of nonlinear equations, but also, as illustrated by step 4 above, it is able to find a nontrivial regime of parameters where the behavior of the network can be significantly improved. Here this is done by reinforcing simultaneously the weight of two branches of the network implicated in a crucial incoherent feedforward loop, implementing perfect adaptation, and allowing to define a simple adaptation submodule. $\tau$ dependency is encoded downstream this adaptation module in $C_1$, defining a kinetic sensing submodule. A general feature of $\bar \phi$ is its ability to identify and reinforce crucial functional parts in the networks, as will be further illustrated below.

\subsection*{$\bar \phi$ for  SHP-1 model}

\begin{figure}
\includegraphics[width=6.5 in]{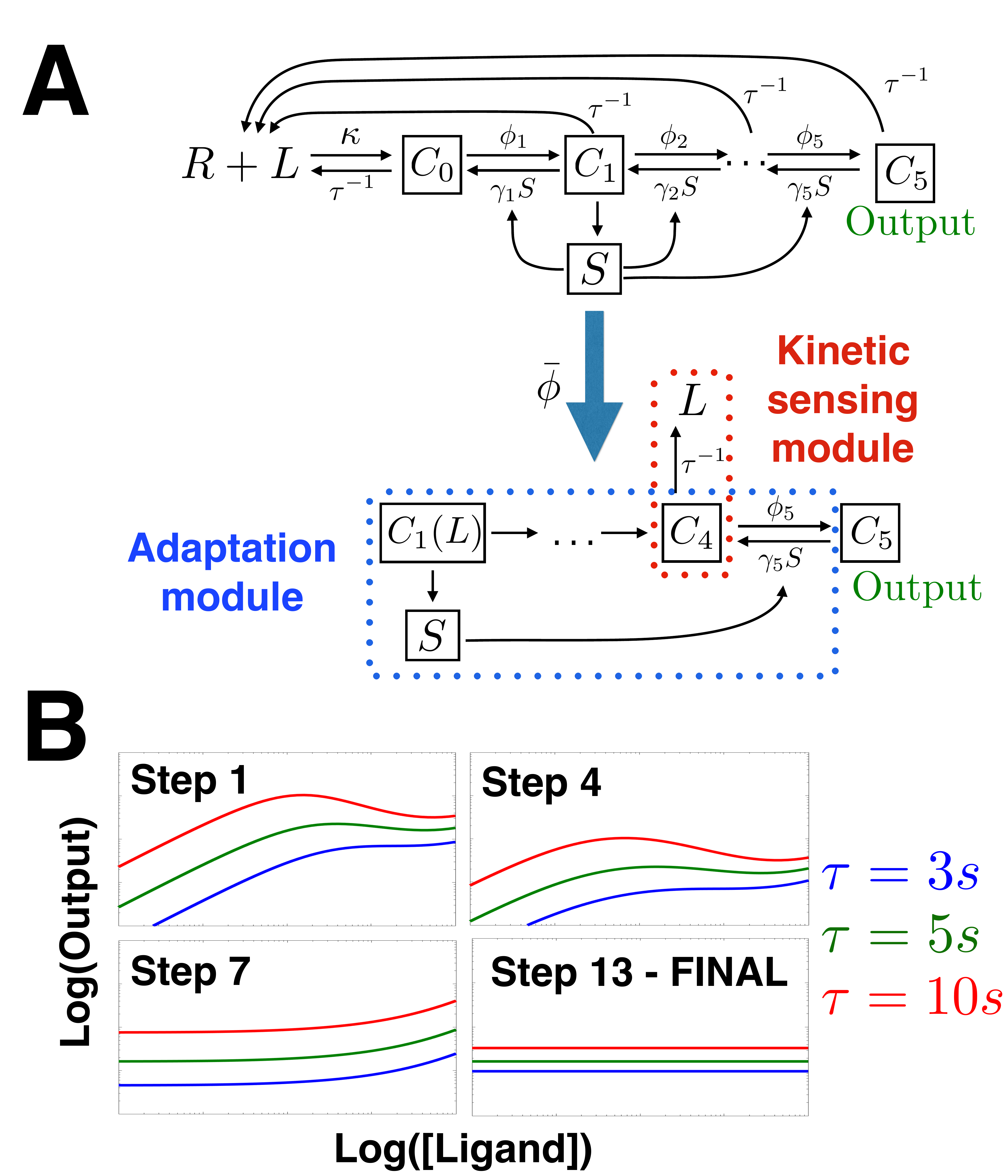}
\caption{ Reduction of SHP-1 model. (A) Initial model considered  and final reduced model (bottom).  Step 1 shows the initial dynamics. Equations can be found in the Supplement.  $\bar \phi$ (with parameter symmetry breaking) eliminates most of the feedback interactions by $S$, separating the full network into an adaptation module and a kinetic sensing module. See main text for discussion. (B) Dose response curves for 3 different values of $\tau=3,5,10 s$ and different steps of $\bar \phi$ reduction, showing how the curves become more and more horizontal for different $\tau$, corresponding to better absolute discrimination. Corresponding parameter modifications are given in the Supplement. FINAL panel shows behavior of Eqs. 9-15 in the Supplement (full system including local mass conservation). }\label{fig-shp}
\end{figure}

This model aims at modelling early immune recognition by T cells  \cite{Francois:2013} and combines a classical proofreading cascade \cite{McKeithan:1995} with a negative feedback loop (Fig. \ref{fig-shp} A, top).  The proofreading cascade amplifies the $\tau$ dependency of the output variable, while the variable $S$ in the negative feedback encodes the ligand concentration $L$ in a nontrivial way. The full network presents dose response-curves plateauing at different values for different $\tau$s, allowing for approximate discrimination as detailed in \cite{Francois:2013}  (Fig. \ref{fig-shp} B, step 1). Full understanding of the steady state requires solving a $N \times N$ linear system in combination with a polynomial equation of order $N-1$, which is analytically possible if $N$ is small enough (See Supplement). Behavior of the system can only be intuitively grasped in limits of strong negative feedback and infinite ligand concentration \cite{Francois:2013}.   The logic of the network appears superficially similar to the previously described adaptive sorting network, with a competition between proofreading and feedback effects compensating for $L$, thus allowing for approximated kinetic discrimination based on parameter $\tau$. Other differences include the sensitivity to ligand antagonism because of the different number of proofreading steps, discussed in \cite{Francois:2016ig} .

When performing $\bar \phi$ on this model, the algorithm quickly gets stuck without further reduction in the number of parameters and corresponding network complexity. By inspection of the results, it appears that the network is too symmetrical: variable $S$ acts in exactly the same way on all proofreading steps at the same time. This creates a strong nonlinear feedback term that explains why the nonmonotonic dose-response curves are approximately flat as $L$ varies as described in \cite{Francois:2013}, as well as other features, such as loss of response at high ligand concentration that is sometimes observed experimentally. This also means the output can never be made fully independent of $L$  (see details in the Supplement). But it could  also be interesting biologically to explore limits where dephosphorylations are more specific, corresponding  to breaking symmetry in parameters .

We thus perform symmetry breaking, so that $\bar \phi$ converges in less than 15 steps, as shown in one example presented in Fig. \ref{fig-shp}. The dose-response curves as functions of $\tau$ become flatter while the algorithm proceeds, until perfect absolute discrimination is reached (flat lines on Fig \ref{fig-shp} B, step 13). 

A summary of the core network extracted by $\bar \phi$ is presented in Fig. \ref{fig-shp} A. In brief, symmetry breaking in parameter space concentrates the functional contribution of $S$ in one single network interaction. This actually  {\it reduces} the strength of the feedback, making it exactly proportional to the concentration of the first complex in the cascade $C_1$, allowing for a better balance between the negative feedback and the input signal in the network.

Eventually,  the dynamics of the last two complexes in the cascade are given by :

\begin{eqnarray}
\dot C_4 &=& \phi_4 C_3 +\gamma_5 S C_5 - (\phi_5+\tau^{-1}) C_4 \qquad \textrm{with}  \qquad C_3 \propto C_1  \label{C_4_Shp} \\
\dot C_5 &=& \phi_5 C_4 - \gamma_5 S C_5  \qquad \textrm{with}  \qquad S \propto C_1 \label{C_2_Shp}
\end{eqnarray}

Now at steady state, $\phi_5 C_4 = \gamma_5 S C_5 $ from Eq. \ref{C_2_Shp} so that those terms cancel out in Eq. \ref{C_4_Shp} and we get that at steady state $C_4=\phi_4 \tau C_3$, with $C_3$  proportional to $C_1$ via $C_2$ in the cascade. Looking back at Eq. \ref{C_2_Shp}, it means that  at steady state both the production and the degradation rates  of $C_5$ are proportional to $C_1$ (respectively via $C_3$  for production and $S$ for degradation) . This is another tug-of-war effect, so that at steady state $C_5$ concentration is independent of $C_1$ and thus from $L$.  However, there is an extra $\tau$ dependency coming from $C_4$ at steady state (Eq. \ref{C_4_Shp}), so that $C_5$ concentration is simply proportional to a power of $\tau$ (see full equations in the Supplement).

Again, $\bar \phi$ identifies and focuses on different parts of the network to perform perfect absolute discrimination.  Symmetry breaking in the parameter spaces allows to decouple identical proofreading steps and effectively makes the behavior of the network more modular, so that only one complex in the cascade is responsible for the $\tau$ dependency  (``kinetic sensing module" in Fig. \ref{fig-shp})  while another one carries the negative interaction of $S$  (``Adaptation module" in Fig. \ref{fig-shp}) .
  
 When varying initial parameters for reduction, we see different possibilities for the reduction of the network (see examples in the Supplement). While different branches for degradation by $S$ can be reinforced by $\bar \phi$,  eventually only one of them  performs perfect adaptation. Similar variability is observed for $\tau$ sensing. Another reduction of this network is presented in the Supplement.

\subsection*{$\bar \phi$  for Lipniacki model}

While the $\bar \phi$ algorithm works nicely on the previous examples, the models are simple enough so that in retrospect the reduction steps might appear as natural (modulo nontrivial effects such as mass conservation or symmetry breaking). It is thus important to validate the approach on a more complex model which can be understood intuitively but is too complex mathematically to assess without simulations, a situation typical in systems biology. It is also important to apply $\bar \phi$ to a published model not designed by ourselves.

\begin{figure}
\includegraphics[width=6 in]{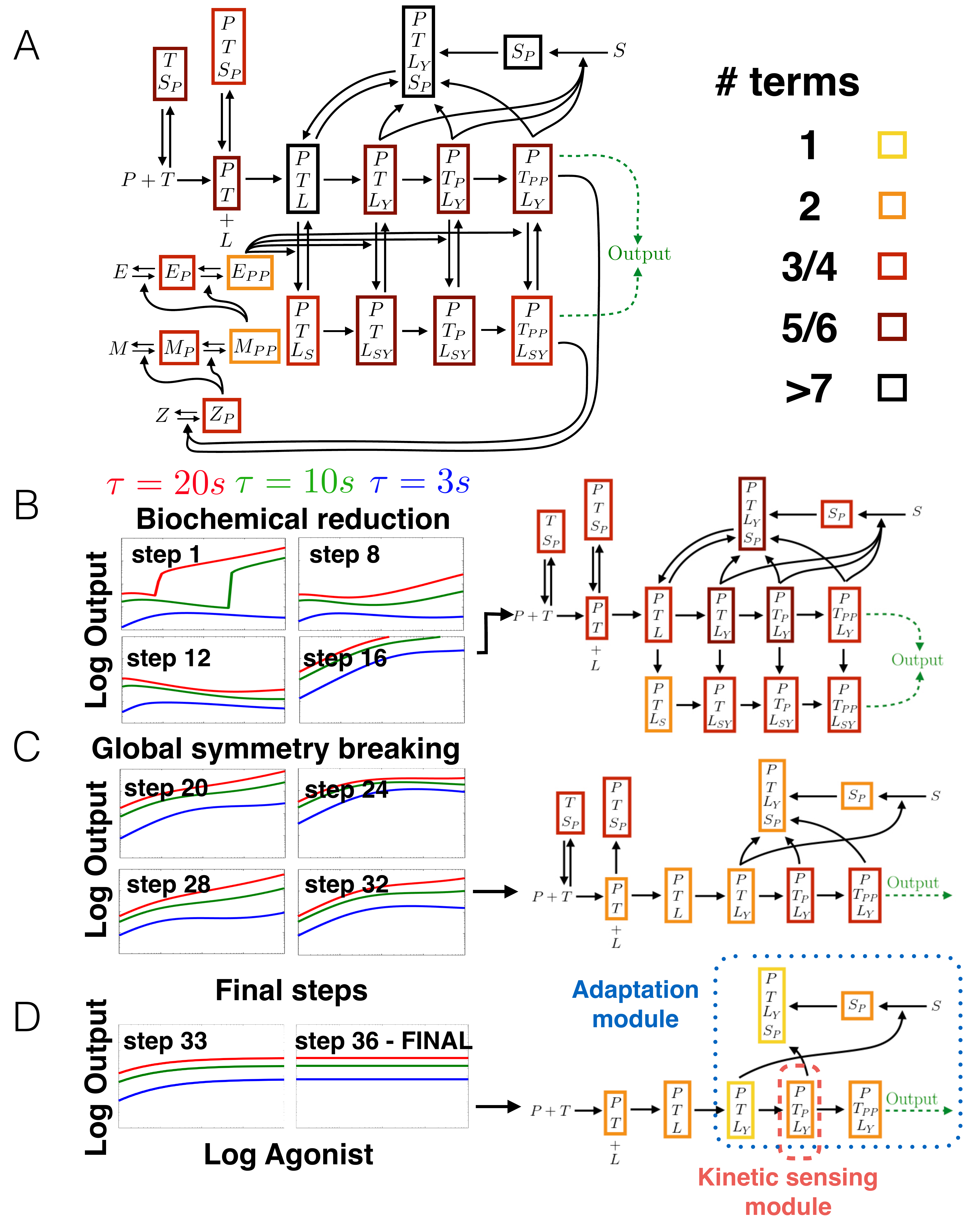}
\caption{ Reduction of Lipniacki model. (A) Initial model considered. We indicate complexity with coloured squared boxes that correspond to the number of individual reaction rates in each of the corresponding differential equations for a given variable. (B)  to (D) Dose response curves for different reduction steps.   Step 1 shows the initial dynamics. From top to bottom, graphs on the right column displays the (reduced) networks at the end of steps 16 (biochemical reduction), 32 (symmetry breaking), 36 (final model). The corresponding parameter reduction steps are given in the Supplement.  FINAL panel shows behavior of  Eq. 28-34 in the Supplement (full system including local mass conservation).  }\label{fig-lip1}
\end{figure}

\begin{figure}
\includegraphics[width=6.5 in]{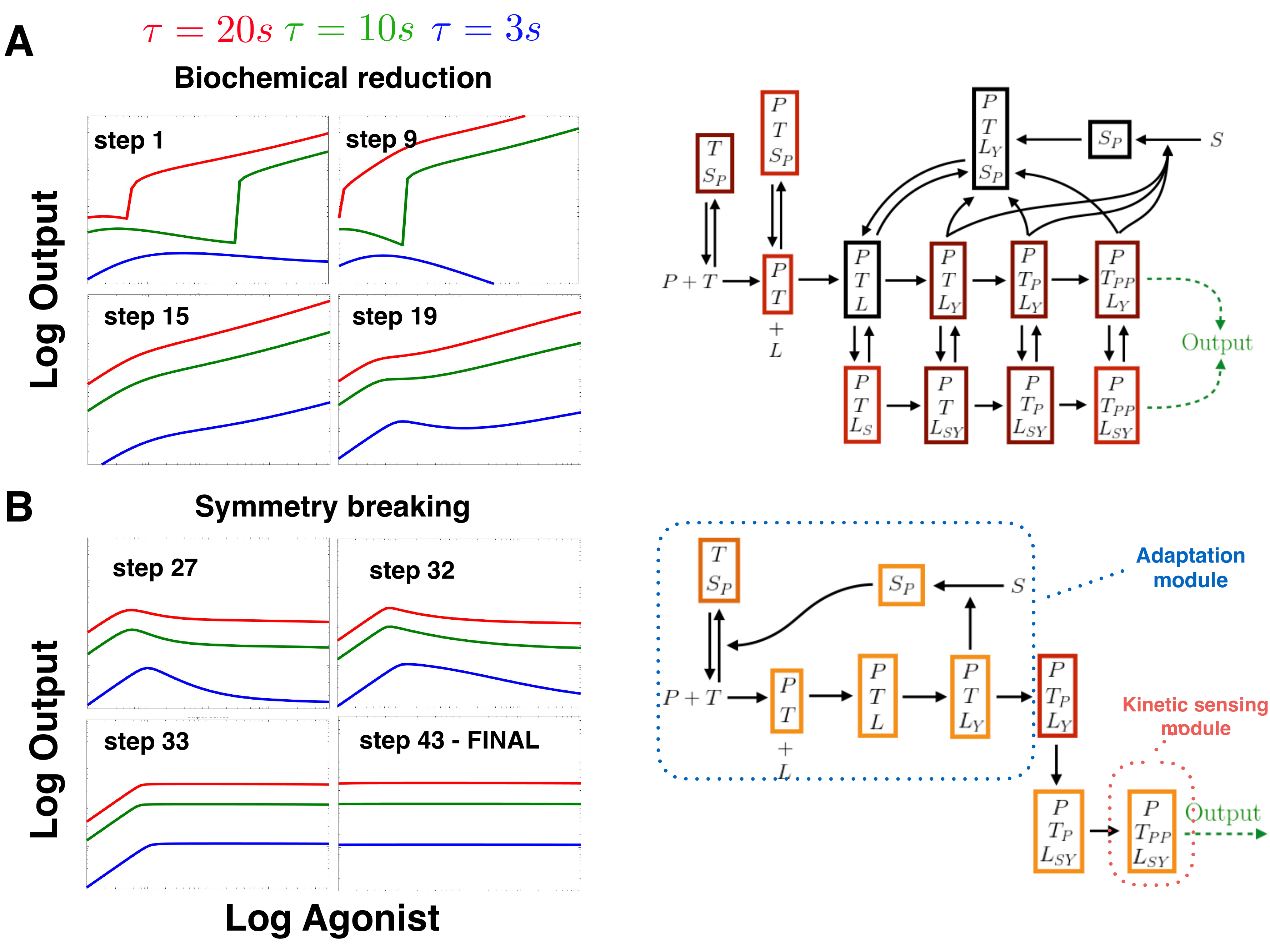}
\caption{Another reduction of the Lipniacki model starting from the same network as in Fig. \ref{fig-lip1} A leading to a different adaptive mechanism. The corresponding parameter reduction steps are given in the Supplement. (A) Initial biochemical reduction suppresses the positive feedback loop in a similar way (compare with Fig. \ref{fig-lip1} B). (B) Symmetry breaking breaks proofreading cascades and isolates different adaptive and kinetic modules (compare   with Fig. \ref{fig-lip1} D). FINAL panel shows behavior of  Eq. 35-43 in the Supplement (full system including local mass conservation). }\label{fig-lip2}
\end{figure}

We thus consider a much more elaborated model for T cell recognition proposed in \cite{Lipniacki:2008} and inspired by \cite{AltanBonnet:2005}. This models aims at describing many known interactions of receptors in a realistic way, and accounts for several kinases such as Lck, ZAP70, ERK, and phosphatases such as SHP-1, multiple phosphorylation states of the internal ITAMs. Furthermore, this model accounts for multimerization of receptors with the enzymes. As a consequence, there is an explosion of the number of cross-interactions and variables in the system, as well as associated parameters (since all enzymes modulate variables differently), which renders its intractable without numerical simulations.  It is nevertheless remarkable that this model is able to predict a realistic response line (e.g. Fig. 3 in \cite{Lipniacki:2008}), but its precise quantitative origin is unclear. The model is specified in the Supplement by its twenty-one equations that include a hundred odd terms corresponding to different biochemical interactions.
With multiple runs of $\bar \phi$ we found two variants of reduction.
 Figs. \ref{fig-lip1} and \ref{fig-lip2} illustrate examples of  those two variants, summarizing the behavior of the network at several reduction steps. Due to the complexity of this network, we first proceed with biochemical reduction. Then we use the reduced network and perform symmetry breaking.

  The network topology at the end of both reductions is shown in Figs. \ref{fig-lip1} and \ref{fig-lip2} with examples of the network for various steps. Interestingly, the steps of the algorithm correspond to successive simplifications of clear biological modules that appear in retrospect unnecessary for absolute discrimination (multiple runs yield qualitatively similar steps of reduction).  In both cases, we observe that biochemical optimization first prunes out the ERK positive feedback  module (which in the full system amplifies response), but keeps many proofreading steps and cross-regulations.  The optimization eventually gets stuck because of the symmetry of the system, just like we observed in the SHP-1 model from the previous section (Fig. \ref{fig-lip1} B and Fig. \ref{fig-lip2} A ).
  
Symmetry breaking is then performed, and allows is to considerably reduce the combinatorial aspects of the system, reducing the number of biochemical species and  fully eliminating one parallel proofreading cascade (Fig. \ref{fig-lip1} C) or combining two cascades (Fig. \ref{fig-lip2} B). In both variants, the final steps of optimization allow for further reduction of the number of variables keeping only one proofreading cascade in combination with a single loop feedback via the same variable (corresponding to phosphorylated SHP-1 in the complete model).

Further study of this feedback loop  reveals that it is responsible for biochemical adaptation, similarly to what we observed in the case of the SHP-1 model.  However, the mechanism for adaptation is different for the two different variants and corresponds to two different parameter regimes.

 For the variant of Fig.  \ref{fig-lip1}, the algorithm converges to a local optimum for the fitness. However upon inspection, the structure appears very close to the SHP-1 model reduction, and can be optimized by putting three additional parameters to $0$.
The Output of the system of Fig. \ref{fig-lip1} is then governed by three variables out of the initial twenty-one and is summarized by: 
\begin{eqnarray}
\dot C_7 &=& \phi_1 C_5(L) - \phi_2 C_7-\gamma S C_7  \label{C-lip}\\ 
\dot S &=& \lambda C_5(L) - \mu R_{tot} S \label{S-lip} \\
\dot C_N &=& \phi_2 C_7- \tau^{-1} C_N \label{C_N-lip}
\end{eqnarray}
Here $C_5(L)$ is one of the complex concentrations  midway of the proofreading cascade (we indicate here $L$ dependency that can be computed by mass conservation but is irrelevant for the understanding of the mechanism). $S$ is the variable accounting for phosphatase SHP-1 in the Lipniacki model, and $R_{tot}$ the total number of unsaturated receptors (the reduced system with the name of the original variables is given in the Supplement).

At steady state  $S$ is proportional to $C_5(L)$ from Eq. \ref{S-lip}. We see from Eq. \ref{C-lip} that the production rate of $C_7$ is also proportional to $C_5(L)$. Its degradation rate  $\phi_2+\gamma S$   is proportional to $S$  if $\phi_2 \ll \gamma S$ (which is the case). So both the production and degradation rates of  $C_7$ are  proportional (similar to  what happens in the SHP-1 model, Eq. \ref{C_2_Shp}), and the overall contribution of $L$ cancels out. This corresponds to an adaptation module.

One $\tau$ dependency remains downstream of $C_7$ through Eq. \ref{C_N-lip} (realizing a kinetic sensing module) so that the steady state concentration of $C_N$ is a pure function of $\tau$ , thus realizing absolute discrimination. Notably, this model corresponds to a parameter regime where most receptors are free from phosphatase SHP-1, which actually allows for the linear relationship between  $S$  and $C_5$.

 For the second variant,  when the system has reached optimal fitness the same feedback loop in the model performs perfect adaptation, and the full system of equations in both reductions have similar structure (compare Eqs. 28 - 34  to  Eqs. 35 - 43 in the Supplement). But the mechanism for adaptation is different: this second reduction corresponds to a regime where receptors are essentially all titrated by SHP-1. More precisely, we have (calling $R_f$ the free receptors, and $R_p$ the receptors titrated by SHP-1):

\begin{eqnarray}
\dot R_p &=& \mu  R_f(L) S-\epsilon R_p  \label{R-lip2} \\ 
\dot S &=& \lambda  C_5 - \mu  R_f (L)S  \label{S-lip2} \\
\dot C_5 &=& C_3(L)-  l S C_5 \label{C-lip2} 
\end{eqnarray}

Now at steady state, $\epsilon$ is small so that almost all receptors are titrated in the form $R_p$, and thus $R_p \simeq R_{tot}$. This fixes the product $ R_f (L) S\propto R_{tot}$ to a value independent of $L$ in Eq. \ref{R-lip2}, so that at steady state of $S$ in Eq. \ref{S-lip2}, $C_5= \epsilon R_{tot}/\lambda $ is itself fixed at a value independent of $L$. This  implements an ``integral feedback'' adaptation scheme \cite{Yi:2000tj}. Down $C_5$, there is a simple linear cascade where one $\tau$ dependency survives, ensuring kinetic sensing and absolute discrimination for the final complex of the cascade.

\subsection*{Comparison and categorization of models}

\begin{figure}
\includegraphics[width=6.5 in]{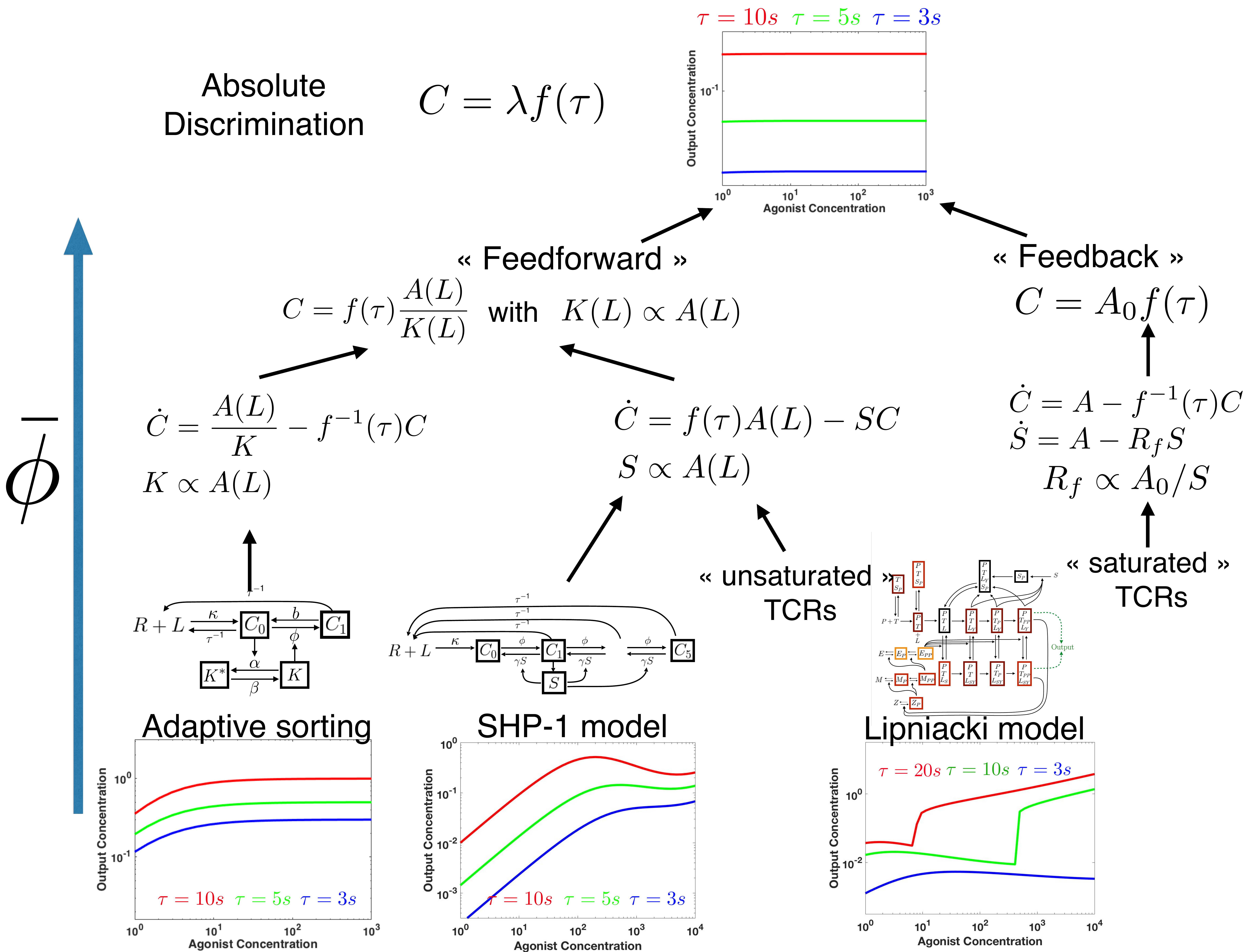}
\caption{ Categorization of networks based on $\bar \phi$ reduction. Absolute discrimination models  considered here (bottom of the tree) can all be coarse-grained into the same functional forms (top of the tree). Intermediate levels in reduction correspond to two different mechanisms, ``feedforward" based and ``feedback'' based. See main text for discussion. }\label{fig-tree}
\end{figure}

An interesting feature of $\bar\phi$ is that reduction allows to formally classify and connect  models of different complexities. We focus here on absolute discrimination only.  Our approach allows us to distinguish at least four levels of coarse-graining for absolute discrimination, as illustrated in Fig. \ref{fig-tree}.

At the upper level, we observe that all reduced absolute discrimination models considered can be broken down into two parts of similar functional relevance. In all reduced models, we can clearly identify an adaptation module realizing perfect adaptation (defining an effective parameter $\lambda$ in Fig. \ref{fig-tree}) , and a kinetic sensing module performing the sensing of $\tau$ (function $f(\tau)$ in Fig. \ref{fig-tree}). If $f(\tau)=\tau$, we get a two-parameter model, where each parameter relates to a submodule.

The models can then be divided in the nature of the adaptatation module, which gives the second level of coarse-graining. With $\bar \phi$, we  automatically recover a dichotomy previously observed for biochemical adaptation between feedforward and feedback models  \cite{Francois:2008,Ma:2009}. The second variant of Lipniacki relies on an integral feedback mechanism, where adaptation of one variable ($C_5$)  is due to the buffering of a negative feedback variable ($S(L)$)  (Eqs. \ref{R-lip2} - \ref{C-lip2}, Fig. \ref{fig-tree}).  Adaptive sorting, the SHP-1 model and the first variant of Lipniacki model instead rely on a ``feedforward'' adaptation module where a tug-of-war between two terms (an activation term $A(L)$ and feedforward terms $K$ / $S$ in Fig. \ref{fig-tree})  exactly compensates.

  The tug-of-war necessary for adaptation is realized in two different ways, which is the third level of coarse-graining. In adaptive sorting, this tug-of-war is realized at the level of the  {\it production} rate of the Output, that is made ligand independent by a competition between a direct positive contribution and an indirect negative one (Eq. \ref{C1}, Fig. \ref{fig-tree}). In the reduced SHP-1 model, the {\it concentration} of the complex $C$ upstream the output is made $L$ independent via a tug-of-war between its production and degradation rates. The exact same effect is observed in the first variant of the Lipniacki model: at steady state, from Eqs. \ref{C-lip} and \ref{S-lip} the production and degradation rates of $C_7$ are proportional  (Fig. \ref{fig-tree}) which ensures adaptation. So $\bar \phi$ allows to  rigorously confirm the intuition that the SHP-1 model and the Lipniacki model indeed work in a similar way and belong to the same category in the unsaturated receptor regime. We also notice that $\bar \phi$ suggests a new coarse-grained model for absolute discrimination based on modulation of degradation rates, with fewer parameters and simpler behavior than the existing ones, by assuming specific dephosphorylation in the cascades (we notice that some other models have suggested specificity for the last step of the cascade, e.g. in limited signalling models \cite{Lever:2014}).
  
 Importantly, the variable $S$, encoding for the same negative feedback in both the SHP-1 and the first reduction of Lipniacki model, plays a similar role in the reduced models, suggesting that two models of the same process, while designed with different assumptions and biochemical details, nevertheless converge to the same class of models. This variable $S$ also is the buffering variable in the integral feedback branch of the reduction of the Lipniacki model, yet adaptation works in a different way for this reduction. This shows that even though the two reductions of the Lipniacki model work in different parameter regimes and rely on different adaptive mechanisms, the same components in the network play the crucial functional roles, suggesting that the approach is general. As a negative control of both the role of SHP-1 and more generally of the $\bar \phi$ algorithm, we show in Supplement on the SHP-1 model that reduction does not converge in the absence of the $S$ variable (Fig. S3).

Coarse-graining further allows us to draw connections between network components and parameters for those different models. For instance, the outputs are functions of $K(L)C_0(L)$ for adaptive sorting and of  $\frac{C(L)}{S(L)}$ for SHP-1/Lipniacki models, where $C_0(L)$ and $C(L)$ are in both models concentrations of complex upstream in the cascade. So we can formally identify $K(L)$ with $S(L)^{-1}$. The immediate interpretation is that deactivating a kinase is  similar to activating a phosphatase, which is intuitive but only formalized here by model reduction.

At lower levels in the reduction, complexity is increased, so that many more models are expected to be connected to the same functional absolute discrimination model. For instance, when we run $\bar \phi$ several times, the kinetic discrimination module on the SHP-1 model is realized on different complexes (see several other examples in the Supplement). Also, the precise nature and position of kinetic discriminations in the network might influence properties that we have not accounted for in the fitness. In the Supplement, we illustrate this on ligand antagonism \cite{Francois:2016kt}:  depending on the complex regulated by $S$ in the different reduced models, and adding back kinetic discrimination (in the form of $\tau^{-1}$ terms) in the remaining cascade on the reduced models, we can observe different antagonistic behaviour, comparable with the experimentally measured antagonism hierarchy (Fig. S4 in Supplement). Finally, a more realistic model might account for nonspecific interactions (relieved here by parameter symmetry breaking), which might only give approximate biochemical adaptation (as in \cite{Francois:2013}) while still keeping the same core principles (adaptation + kinetic discrimination) that are uncovered by $\bar \phi$.

\section*{Discussion}


When we take into account all possible reactions and proteins in a biological network, a potentially infinite number of different models can be generated. But it is not clear how the level of complexity relates to the behavior of a system, nor how models of different complexities can be grasped or compared. For instance, it is far from obvious whether a network as complex as the one from \cite{Lipniacki:2008} (Fig. \ref{fig-lip1} A)  can be simply understood in any way, or if any clear design principle can be extracted from it. We propose $\bar \phi$, a simple procedure to reduce complex networks, which is based on a fitness function that defines network phenotype, and on simple coordinated parameter changes.

 $\bar \phi$ relies on the optimization of a predefined fitness that is required to encode coarse-grained phenotypes. It performs a direct exploration of the asymptotic limit on boundary manifolds in parameter space.   {\it In silico} evolution of networks teaches us that the choice of fitness is crucial for successful exploration in parameter spaces and to allow for the identification of design principles \cite{Francois:2014}.  Fitness should capture qualitative features of networks that can be improved  incrementally; an example used here is mutual information \cite{ Lalanne:2013}.  While adjusting existing parameters or even adding new ones (potentially leading to overfitting) could help optimizing this fitness, it is not obvious  \textit {a priori} that systematic \textit {removal} of parameters is possible without decreasing the fitness, even for networks with initial good fitness. For both cases of biochemical adaptation and absolute discrimination, $\bar \phi$ is nevertheless efficient at pruning and reinforcing different network interactions in a coordinated way while keeping an optimum fitness, finding simple limits in network space, with submodules that are easy to interpret. Reproducibility in the simplifications of the networks suggests that the method is robust.

In the examples of SHP-1 and Lipniacki models, we notice that $\bar \phi$ disentangles the behavior of a complex network into two submodules with well identified functions, one in charge of adaptation and the other of kinetic discrimination.  To do so, $\bar \phi$ is able to identify and reinforce tug-of-war terms, with direct biological interpretation. This allows for a formal comparison of models. The reduced SHP-1 model and the first reduction of the Lipniacki model have  a similar feedforward structure, controlled by a variable corresponding to phosphatase SHP-1 defining the same biological interaction. This is reassuring since both models aim to describe early immune recognition; this was not obvious \textit{a priori} from the complete system of equations or the considered network topology (compare Fig. \ref{fig-shp} with Fig. \ref{fig-lip1}A). These feedforward dynamics discovered by $\bar \phi$  contrast with the original feedback interpretation of the role of SHP-1 from the network topology  only  \cite{AltanBonnet:2005,Lipniacki:2008,Francois:2013}. Adaptive sorting, while performing the same biochemical function, works  differently by adapting the production rate of the output, and thus belongs to another category of networks  (Fig. \ref{fig-tree}).

$\bar \phi$ is also able to identify different parameter regimes for a network performing the same function, thereby uncovering an unexpected network plasticity. The two reductions of the Lipniacki model work in a different way (one is feedforward based, the other one is feedback based), but importantly, the crucial adaptation mechanism relies on the same node, again corresponding to phosphatase SHP-1, suggesting the predictive power of this approach irrespective of the details of the model. From a biological standpoint, since  the same network can yield two different adaptive mechanisms depending on the parameter regime (receptors titrated or not by SHP-1), it could be that both situations are observed. In mouse,  T Cell Receptors (TCRs) do not bind to phosphatase SHP-1 without engagement of ligands \cite{Dittel:1999}, which would be in line with the reduction of the SHP-1 model and the first variant of the Lipniacki model reduction. But we cannot exclude that a titrated regime for receptors exists, e.g. due to phenotypic plasticity \cite{Feinerman:2008a}, or that the very same network works in this regime in another organism. More generally, one may wonder if the parameters found by $\bar \phi$ are realistic in any way. In cases studied here, the values of parameters are not as important as the regime in which the networks behave. For instance, we saw for the feedforward models that some specific variables have to be proportional, which requires nonsaturating enzymatic reactions. Conversely, the second reduction of the Lipniacki model requires titration of receptors by SHP-1. These are direct predictions on the dynamics of the networks, not specifically tied to the original models.

Since $\bar \phi$ works by sequential modifications of parameters,  we get a continuous mapping between all the models at different steps of the reduction process, via the most simplified one-parameter version of the model.  By analogy with physics, $\bar \phi$ thus ``renormalizes" different networks by coarse-graining  \cite{Machta:2013ga}, possibly identifying universal classes for a given biochemical computation, and defining subclasses \cite{Transtrum:tq7xs8Lb}.  This allows  us to  draw correspondences between networks with very different topologies, formalizing ideas such as the equivalence between activation of a phosphatase and repression of a kinase (as exemplified here by the comparison of influences of $K(L)$ and $S(L)$ in reduced models from Fig. \ref{fig-tree}). In systems biology, models are neither traditionally simplified, nor are there systematic comparisons between models, in part because there is no obvious strategy to do so. The approach proposed here offers a solution for both comparison and reduction, which complements other strategies such as  the evolution of phenotypic models \cite{Francois:2014} or direct geometric modelling in phase space \cite{Corson:2012}.

To fully reduce complex biochemical models, we have to perform symmetry breaking on parameters.  Similar to parameter modifications,  the main roles of symmetry breaking is to reinforce and adjust dynamical regimes in different branches of the network, e.g. imposing proportionality to tug-of-war terms. Intuitively, symmetry breaking embeds complex networks into a higher dimensional parameter space allowing for better optimization.   Much simpler networks can be obtained with this procedure, which shows in retrospect how the assumed nonspecificity  in interactions strongly constrains the allowed behavior. Of course, in biology, some of this complexity might also have evolutionary adaptive values, corresponding to other phenotypic features we have neglected here, such as signal amplification. A tool like $\bar \phi$ allows for a reductionist study of these features by specifically focusing on one phenotype of interest to extract its core working principles. Once the core principles are identified, it should be easier to complexify a model by accounting for other potential adaptive phenotypes (e.g. as is done to reduce antagonism in \cite{ Lalanne:2013} or in Supplement in Fig. S4) .

Finally, there is a natural evolutionary interpretation of $\bar \phi$. In both evolutionary computations and evolution, random parameter modifications in evolution can push single parameters to $0$ or potentially very big values (corresponding to the $\infty$ limit). However, it is clear from our simulations that concerted modifications of parameters are needed, e.g. for adaptive sorting, the simultaneous modifications of the kinetics and the efficiency of a kinase regulation is required  in Step 4 of the reduction. Evolution might select for networks explicitly coupling parameters that need to be modified in concert. Conversely, there might be other constraints preventing efficient optimizations in two  directions in parameter space at the same time, due to epistatic effects. Gene duplications provide an evolutionary solution to relieve such trade-offs, after which previously identical genes can diverge and specialize \cite{Innan:2010gf}. This clearly bears resemblance to the symmetry breaking proposed here. For instance, having two duplicated kinases instead of one would allow to have different phosphorylation rates in the same proofreading cascades. We also see in the examples of Figs. \ref{fig-shp}, \ref{fig-lip1}, and \ref{fig-lip2}  that complex networks that cannot be simplified by pure parameter changes, can be improved by parameter symmetry breaking via decomposition into independent submodules. Similar evolutionary forces might be at play to explain the observed modularity of gene networks \cite{Milo:2002}. More practically, $\bar \phi$ could be useful as a complementary tool for artificial or simulated evolution \cite{Francois:2014} to simplify complex simulated dynamics \cite{Sussillo:2009vd}.

\section*{Author contributions}

Conceptualization: PF; Methodology:  TR and PF; Software,Validation, Formal Analysis,  Investigation: all 3 authors; Writing: PF; Funding Acquisition: FPG and PF; Project Administration, Supervision: PF.

\section*{Acknowledgments}

We thank the members of the Fran\c cois group for their comments on the manuscript.  We also thank anonymous referees for useful comments.


%
%
%

\end{document}


\title{Supplement}

\maketitle 

\section{Description of $\bar \phi$}

In the following, we illustrate $\bar \phi$'s MATLAB implementation. The MATLAB code is available in the Supplementary Materials.

\subsection{MATLAB Algorithm}
\label{sect:algo}
{sect:algorithm}
Seven functions are used: \\
runLoop \\
paramMODELtype \\
calcFitness \\
odeMODELtype \\
evalLim \\
updateParam \\ 
catch\_problems \\ \\
runLoop is the main script. paramMODELtype  and odeMODELtype define  parameters and associated ODEs which are problem specific. Parameters associated to the model are initially stored in variable \textit{default}, then later modified parameters are stored in variable \textit{param} and the list of removed parameters is stored in variable \textit{removed}

\begin{figure}
\includegraphics[width=\textwidth]{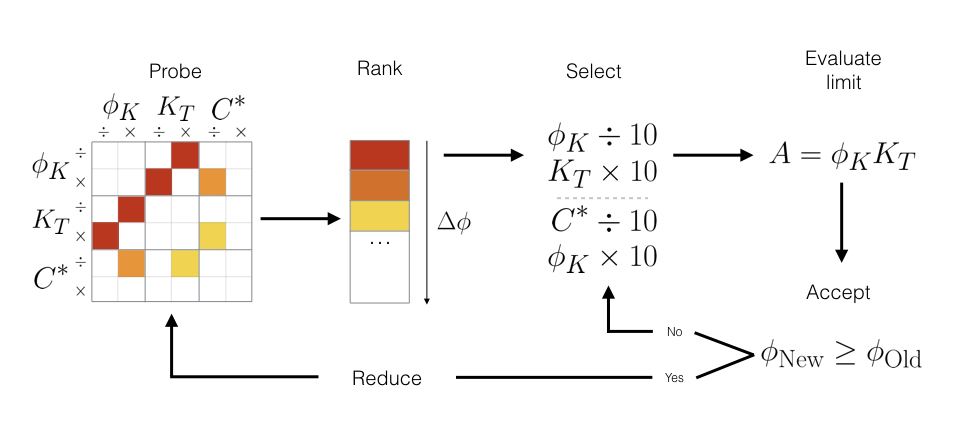}
\centering
\caption{A flowchart of the algorithm.}
\label{fig:flowchart}
\end{figure}

A flowchart of the algorithm is presented in Fig. S\ref{fig:flowchart}. In the following five steps we probe (1 - 3), rank, select, evaluate, accept (4), reduce and repeat (5).  \\

1. Assign the parameter vector (PV) that paramMODELtype returns to \textit{default.} This point in parameter space is going to be probed. \\

2. The fitness landscape around the initial PV \textit{default} is characterized by the symmetric matrix \textit{fitnessmap}, containing the fitness for modified parameters or couples of parameters.  The fitness function, on the contrary, is problem specific, and is computed by calcFitness. Row by row, \textit{fitnessmap} is filled by multiplying/dividing the parameters per entry by a reacaling \textit{factor} ($f=10$ from the main text). The performance of each of these entries is measured by computing the fitness with the new parameter combination relative to the initial fitness. A network with $N$ parameters has $2N^2$ independent entries. \\

3. Removing a parameter is done in evalLim. With an estimate of the fitness landscape at hand found via the previous steps, the algorithm takes the corresponding limit (to $0$ or $\infty$) for the parameters that were rescaled by $f$. We consider only changes of parameters giving identical or improved fitness. There exist four groups of two parameter limits $\theta_i, \, \theta_j$. In Tab. \ref{tab:limits}, the groups are presented in order of importance. When several couples of parameters give favorable changes to the fitness, we evaluate the limit of all couples that fall in group 1 one by one. \\

4. When we encounter a parameter limit in which the fitness is improved, we eliminate corresponding parameters and return to step 1. If for none of the couples in the parameter limits the fitness is improved, we move to the members of group 2, the limits to infinity, and similarly when we find a parameter limit that improves the fitness, we reduce and move on. Otherwise  we move to the parameter limits of groups 3 and, finally to group 4 with the same criteria. This natural order shows our preference for removing parameters one by one (set parameters values to zero), instead of simply rescaling them (as products). Notice that we take a very conservative approach where fitness can only be incrementally improved with this procedure.


The steps in evalLim are the following.
\\

\begin{enumerate}[label=\Alph*]
\item Find the least nonnegative elements in \textit{fitnessmap}
\item Divide these in the groups defined above
\item Pick a random element from the highest ranked nonempty group
\item updateParam takes the PV \textit{default} and a $2 \times 2$ block of \textit{removed} as arguments and returns an updated PV to \textit{param.}
\item Compute a temporary fitness $\phi_{\text{new}}$ with \textit{param.}
\item Decide as follows: \\
If $\phi_{\text{new}} \geq \phi_{\text{init}}.$ \\
\phantom{.} 	\qquad Accept removal \\
\phantom{.} 	\qquad Return \textit{param} and \textit{removed} \\
If $\phi_{\text{new}} < \phi_{\text{init}}.$ \\
\phantom{.} 	\qquad Reject removal \\ 
\phantom{.}    \qquad Set \textit{fitnessmap}(picked element) to $\inf.$ \\
\phantom{.} 	\qquad Repeat cycle at step A \\
\end{enumerate}

The method we use to take asymptotic limits is described in the next section.

5. The returned PV becomes the new initial point in an $(N-1)$-dimensional plane that is embedded in $N$-dimensional parameter space. Around this new initial point, we will probe the fitness landscape in the next round.  In \textit{removed}, the removed parameters and their limits are stored such that $\bar{\phi}$ ignores directions of reduced parameters in subsequent rounds. 

This procedure is repeated until there are no free parameters left, or until all directions will decrease the fitness.

\begin{table}[]
\centering
\caption{Four groups of two-parameter limits}
\label{tab:limits}
\begin{tabular}{c|l|l}
Group & Operation               & Corresponding Limit  taken                                           \\ \hline
1     & Division   of two parameters by $f$         & $ (\theta_i, \theta_j) \to 0 $           \\
2     & Multiplication   of two parameters by $f$          & $ (\theta_i, \theta_j) \to \infty$ \\
3     & Division/multiplication by $f$  & $\theta_i \to 0, \theta_j \to \infty$        \\
4     & Division/multiplication by $f$ &     Rescaling keeping product $\theta_i \cdot \theta_j =$ constant  
\end{tabular}
\end{table}

\subsection{Taking asymptotic limits}
\label{sect:lim_to_inf}

There are two kinds of asymptotic limits: parameters are either taken to $0$ or to $\infty$. The $0$ case is trivial to deal with: when a parameter is chosen to be $0$, we simply put and maintain it to $0$ in the subsequent steps of $\bar \phi$.

In evaluating a limit to infinity, one cannot simply numerically set this parameter to infinity, like in the case of a zero-limit. Instead, we consider a limit where this parameter is increased to such an extend that it dominates other terms in a sum that affect the same variable; these other terms are then removed from the equations. More precisely, consider the following equation:
\begin{equation}
\label{eq:ex_infty_lim}
\dot{y}_2 = a y_1 - (b + c + d y_1) y_2.
\end{equation}
In the limit of $b \to \infty$ we replace this equation by the following  differential equation:
\begin{equation}
\label{eq:ex_simpli}
\dot{y}_2 = a y_1 - b y_2, 
\end{equation}
where $b \to b' = f b$, where $f$ is our multiplicative factor defined in the previous section. This  implements the idea that the $c$ and $d y_1$ terms are negligible compared to $b$.

It is important to define a vector of parameter coefficients to keep track of these infinities. The vector of coefficients is attached to the parameter vector and updated in updateParam similarly. When the limit of a parameter is taken to infinity, its coefficient becomes zero, and the other terms in the sum will disappear. Practically, Eq. \ref{eq:ex_infty_lim} is rewritten as 
\begin{equation}
\label{eq:ex_coeff}
\dot{y}_2 = c_d a y_1 - (c_c c_d b + c_b c_d c + c_a c_b c_c d y_1) y_2.
\end{equation}
The coefficients $c_{a,b,c,d}$ are initially set to 1. After evaluating the limit of $b \to \infty,$ we set $c_b = 0,$ and the simplification from Eq. \ref{eq:ex_infty_lim} to Eq. \ref{eq:ex_simpli} indeed takes place.  

This can however create mass conservation problems in the rate equations. Consider the following equations for $\dot{y}_4$ and $\dot{y}_5$  where $y_4$ is turned into $y_5$ with rate $r$

\begin{equation}
\begin{split}
\dot{y}_4 &= a y_3 - (r + q) y_4 \\
\dot{y}_5 &= r y_4 - d y_5 
\end{split}
\end{equation}

In the limit where parameter $q \to \infty$, parameter $r$ will disappear from the equation of $\dot{y}_4$ potentially creating a divergence in the equations. A way to circumvent this is to impose global mass conservation: situations where $y_4$ is turned into $y_5$ correspond to signalling cascades where complexes are transformed into one another, so that we can impose that the total quantity of complex is conserved. This effectively adds a compensating term to the cascade. We also explicitly control for divergences and discard parameter sets for which variables diverge.

\subsection{Choice of the path in parameter space}
As shown in Section \ref{sect:algo}, the matrix \textit{fitnessmap} is analyzed in the function evalLim. This matrix is symmetrical since the upper triangular part of the matrix corresponding to parameters $ (k_1,k_2)$ and the lower triangular part corresponding to parameters $ (k_2,k_1)$ give similar limits for groups 2 and 4 in Tab. \ref{tab:limits}. When given the choice between sending $ (k_1,k_2) \to \infty$ or $ (k_2,k_1) \to \infty$, FIBAR chooses randomly between the two, because the parameter combinations have the same change in fitness and in both cases a new parameter $ k_1/k_2$ can be identified. However, because of FIBAR's design, choosing one will result in a different exploration of parameter space in the remaining steps. By choosing the first parameter combination, $\bar{\phi}$ will effectively freeze $k_1$ but allows $\bar{\phi}$ to keep exploring the logarithmic neighborhood of $k_2$. If the second combination is chosen, then the value of $k_2$ is frozen and it is the neighborhood of $k_1$ that will be probed. $k_2$ and $k_1$ may be present in different equations in the model, resulting in two not necessarily converging reductions.

A choice thus needs to be made in the final parameter reduced model. This allows for introduction of some kind of stochasticity in the produced networks in order to identify recurring patterns in the reduction. It can be a challenge in terms of reproducibility. One way to solve this problem is to set a fixed rule in the function evalLim (using variable seed) which is called the deterministic method in the main article. The method of choice (random or deterministic) is left at the discretion of the user. We indeed see differences in the way networks are reduced, but the final structure of the reduced networks in all these cases can easily be mapped onto one another as described in the main text.

\section{Mathematical definitions of fitness}

In this section, we give mathematical definitions of the fitness functions used for both problems presented in the main text

\subsection{Biochemical adaptation}

For biochemical adaptation, we use a fitness  very similar to the one proposed in \cite{Francois:2008}.

We take as an initial input $I=0.5$, then integrate for $1000$ units of time and shift $I$ to $2$. After waiting another $1000$ time steps, we measure $\Delta O_{ss}$ and $\Delta O_{max}$ as indicated in Figure 2 in the main text, and we take as a fitness $\Delta O_{max}+\frac{\epsilon}{\Delta O_{ss}}$ with $\epsilon=0.01$ for the NFBL model and $\epsilon = 0.001$ for the first variant of the reduction of the IFFL model and $\epsilon = 0.01$ for the second variant.

\subsection{Absolute discrimination}

For absolute discrimination, we use a mutual information score, very similar to the one proposed in \cite{Lalanne:2013}.

Imagine that some ligands with concentration $L$ and binding time $\tau$ are presented to a cell. For ligand concentrations chosen from a specified distribution, we can now compute the typical output distribution $p_{\tau}(O)$ for a given $\tau$. We consider log-uniformly distributed Input concentrations, integrate the system of equations, and generate a histogram of output values $O$ corresponding to the input $L$. We take this as our approximation of $p_{\tau}(O)$.
 
\begin{figure}
\includegraphics[width=\textwidth]{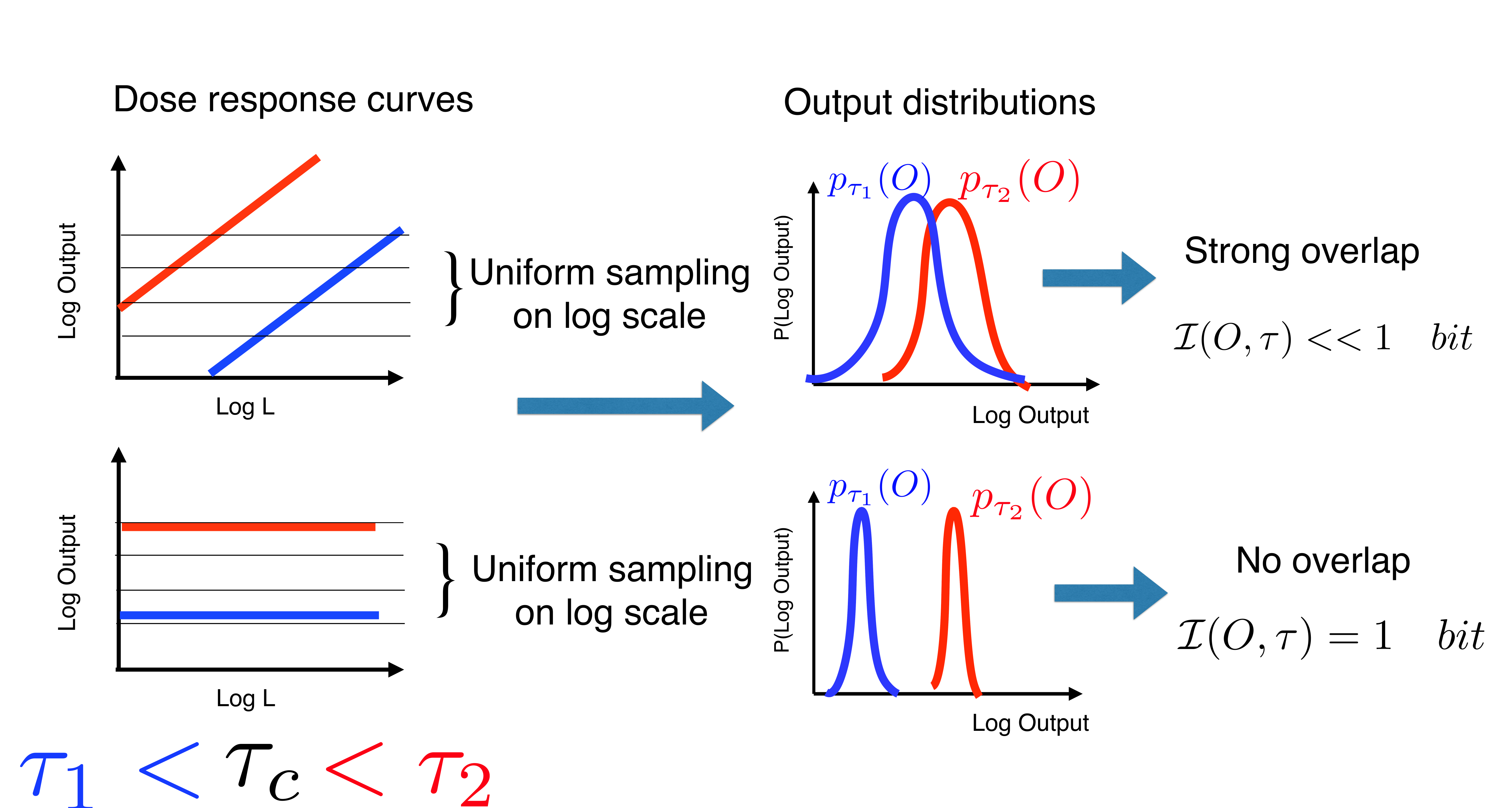}
\caption{Illustration of the mutual information score used for absolute discrimination. Upon sampling $L$ from a log-uniform distribution, we integrate the equations, then compute histograms $p_{\tau_i}(O)$ for $\tau_1$ and $\tau_2$. From those joint distributions we compute the mutual information $\mathcal{I}(O,\tau)$. If the distribution overlap, it is not possible to distinguish well between the two $\tau_i$, which means $\mathcal{I}(O,\tau)$ is low. If the distributions do not overlap, we can unambigously determine $\tau$ given $O$, thus $\mathcal{I}(O,\tau) = 1 \> bit$. }
\label{fig:fitness}
\end{figure}

Absolute discrimination works when the typical output values $O$ across the range of $L$ are unique for different $\tau$. Intuitively, this means that the output distributions for two binding times should overlap as little as possible, as illustrated in Fig. \ref{fig:fitness}. We use these distributions for two values of $\tau$ to define marginal probability distributions $p_{\tau_i}(O)=p(O|\tau_i)$. Lastly, we consider equiprobable $\tau_i$s, and define $p(\tau_i,O)=p(O|\tau_i)p(\tau_i)=\frac{1}{2}p_{\tau_i}(O)$ and compute the mutual information between $O$ and $\tau$ as
\begin{equation}
\mathcal{I}(O,\tau)= H(O)+H(\tau)-H(O,\tau)
\end{equation}
where $H$ is the classical Shannon entropy   $H(O,\tau)=- \sum_{i,O} p(\tau_i,O) \log p(\tau_i,O)$.
 
Mutual information measures how much information we can recover from one variable knowing the other. For instance, when $ \mathcal{I}(O,\tau)=0$, it means we cannot recover information on the value of $\tau$ by observing $O$, which would be the case when both distributions are equal $p_{\tau_1}(O)=p_{\tau_2}(O)$. Conversely, when the two distributions are fully separated, this means we can fully recover $\tau$ by observing $O$. Thus, the the mutual information is at its maximum of 1 $bit$. For partially overlapping distributions, the mutual information varies gradually between $0$ to $1$. The choice of mutual information allows us to focus only on the respective positions of the distributions, and not on their shape, average values, etc... This allows us to focus on the discriminatory phenotype irrespective of other parameters. We very often obtain peaked distribution in $O$ corresponding to horizontal lines (as in the lower panel of Fig. \ref{fig:fitness}); the absolute level of these lines is arbitrary.

During the reduction, we typically sampled 50 log-uniformly distributed $L$ on the interval $[1,10^4]$ and binned the resulting outputs $O$ in 40 log-uniformly distributed bins in the range $[10^{-2}, 10^{2}]$. The results are largely independent from the number of bins or the range of the bins, as long as $O$ remains in the neighborhood of biologically feasible values, the working range of the initial networks. Partly due to this loose constraint, the output of the reduced networks was near the output of the initial networks.

\section{Reductions of biochemical adaptation}

In order to reproduce Michaelis-Menten kinetics for the Negative Feedback Loop and the Incoherent Feedforward Loop models from \cite{Ma:2009}, we consider a model of linear ODEs where the dynamics of a node are mediated by an intermediary enzyme. In the models from \cite{Ma:2009}, the production rates are of the form

\begin{equation}
rate_X = \frac{1-X}{K + 1- X}
\end{equation}

The problem with this is that this imposes $X<1$, which is not enforced by construction in \cite{Ma:2009}. To circumvent this difficulty, we slightly change their model by introducing a production node $ P_X$ and degradation node $ D_X$ to each node $X$. Both $P_X$ and $D_X$ are degraded by $ X$. 
The quasi-static value for the production  ($P_X^{eq}$) and degradation  ($D_X^{eq}$)  nodes are:
\begin{equation}
 P_X^{eq} = \frac{K}{K + X}, \, \, \, \, D_X^{eq} = \frac{K}{K + X}
\end{equation}

so that the full equation for $X$ is given by

\begin{equation}
\dot X = k_p P_X-k_d D_X X= k_p \frac{K}{K + X} - k_d X \frac{K}{K + X}
\end{equation}

with $k_d$ and $k_p$ rates, which are potentially modulated by other proteins. Notice that we have included linear $X$ dependency in the degradation. In particular, addition of $P_X$ ensures that the production rate is a decreasing function of $X$, as was  hypothesized in \cite{Ma:2009}.

\section{Negative feedback loop}

Initial equations for the negative feedback loop model are
given by

\begin{align*}
 \dot{A} &=  k_1 I \, P_A - k_2 A B D_A\\
\dot{P}_A &=  k_3 (1 - P_A) - k_4 A P_A \\
 \dot{D}_A &=  k_5 (1 - D_A) - k_6 A D_A \\
\dot{B} &=  k_7 A \, P_B - k_8 B  D_B\\
 \dot{P}_B &=  k_9 (1 - P_B) - k_{10} B P_B \\
 \dot{D}_B &=  k_{11} (1 - D_B) - k_{12} B D_B. \\
\end{align*}

Initial parameters are given in Tab. \ref{tab:param_NFBL} and steps of the reduction of this model are given in Tab. \ref{tab:details_NFBL}.

\begin{table}[h]
\centering
\caption{Negative feedback loop initial parameters}
\label{tab:param_NFBL}
\begin{tabular}{c|c}
Parameter & Value \\ \hline
$ k_1$ & 2 \\
$ k_2$ & 1 \\
$ k_3$ & 2 \\
$ k_4$ & 1 \\
$ k_5$ & 2 \\
$ k_6$ & 2 \\
$ k_7$ & 2 \\
$ k_8$ & 1 \\
$ k_9$ & 1 \\
$ k_{10}$ & 2 \\
$ k_{11}$ & 1 \\
$ k_{12}$ & 2 \\
\end{tabular}
\end{table}

\begin{table}[h]
\centering
\caption{Negative feedback loop reduction}
\label{tab:details_NFBL}
\begin{tabular}{c|c|c c c|l}
Step & $\phi_{ init}$ & Parameters        &       & Limit         & Description per group   \\ \hline
1       & 1.6794            & $ (k_{12}, k_8)$ & $\to$ & $\infty$      & $ D_B = k_{11}/k_{12}B$  \\ 
2	 & 1.6951            & $ (k_{11}, k_8)$     & $\to$ & $(0, \infty)$ &  $ \Gamma_1 = k_7k_{11}/k_{12}$  \\
3	 & 1.6957            & $ (k_{4},k_9) $     & $\to$ & $ 0,\infty$ & $ P_A = 1$ and $ P_B = 1$ \\ \hline
4	 & 54.4743          & $ k_{10} $    & $\to$ & $ 0 $      &      \\ 
5	 & 54.4743          & $ (k_5,k_6) $    & $\to$ & $\infty $      &      \\  
6	 & 54.5648  	  & $ k_{3}$  & $\to$ & $\infty$ & \\ 
7	 & 54.6099          & $(k_1, k_{2})$		  & $\to$ & $\infty$   &   \\ \hline
\multicolumn{5}{r|}{FINAL OUTPUT} & $ A|_{B_{eq}} =  \frac{\Gamma_1}{k_7} = \frac{k_8k_{11}}{k_7k_{12}}$
\end{tabular}
\end{table}

%
%
%

\section{Incoherent feedforward foop}

For this model, we got two kinds of reductions  and present two variants. Initial equations for the incoherent feedforward loop model are
\begin{align*}
 \dot{A} &=  k_1 I \, P_A - k_2 A  D_A\\
 \dot{P}_A &=  k_3 (1 - P_A) - k_4 A P_A \\
 \dot{D}_A &=  k_5 (1 - D_A) - k_6 A D_A \\
 \dot{B} &=  k_7 A \, P_B - k_8 B  D_B\\
 \dot{P}_B &=  k_9 (1 - P_B) - k_{10} B P_B \\
 \dot{D}_B &=  k_{11} (1 - D_B) - k_{12} B D_B \\
 \dot{C} &=  k_{13} A \, P_C - k_{14} C B D_C\\
 \dot{P}_C &=  k_{15} (1 - P_C) - k_{16} C P_C \\
 \dot{D}_C &=  k_{17} (1 - D_C) - k_{18} C D_C\\
\end{align*}

Initial parameters for the first variant of this model are given in Tab. \ref{tab:param_IFFL1}. Steps and equations of the first variant of the reduction of this model are given in Tab. \ref{tab:details_IFFL1}. Initial parameters for the second variant of this model are given in Tab. \ref{tab:param_IFFL2}. 

\begin{table}[]
\centering
\caption{Incoherent feedforward loop first variant initial parameters}
\label{tab:param_IFFL1}
\begin{tabular}{c|c}
Parameter & Value \\ \hline
$ k_1$ & 2 \\
$ k_2$ & 0.5 \\
$ k_3$ & 3 \\
$ k_4$ & 1 \\
$ k_5$ & 2 \\
$ k_6$ & 3 \\
$ k_7$ & 2 \\
$ k_8$ & 1 \\
$ k_9$ & 3 \\
$ k_{10}$ & 1 \\
$ k_{11}$ & 2 \\
$ k_{12}$ & 1 \\
$ k_{13}$ & 2 \\
$ k_{14}$ & 2 \\
$ k_{15}$ & 3 \\
$ k_{16}$ & 2 \\
$ k_{17}$ & 1 \\
$ k_{18}$ & 3 \\
\end{tabular}
\end{table}

\begin{table}[h!]
\centering
\caption{Incoherent feedforward loop first variant}
\label{tab:details_IFFL1}
\begin{tabular}{c|c|c c c|l}
Step & $\phi_{ init}$ & Parameters        &       & Limit         & Description per group   \\ \hline
1       & 0.5958           & $ (k_{14}, k_{13})$ 	& $\to$ & $\infty$       &    \\ 
2	 & 0.5960           & $ (k_{6},k_{5})$    & $\to$ & $\infty$ &   \\ 
3	 & 0.5961           & $ (k_{15},k_{16}) $   & $\to$ & $\infty$ &    \\
4	 & 0.5962           & $ (k_{10},k_{9})$   & $\to$ & $\infty$     &      \\ 
5	 & 0.5971   	 & $ (k_{13},k_{18})$  & $\to$ & $(0,\infty)$ &  $ D_C = k_{17}/k_{18}C$   \\
6	 & 0.6163	          & $(k_{11}, k_{8})$		  & $\to$ & $(0,\infty)$    &   $ D_B = k_{11}/k_{12}B$ \\ 
7	 & 0.6169		  & $(k_{18},k_{17})$               & $\to$ &  $\infty$     &    \\ 
8      	 & 0.6177		 &	$(k_5,k_2)$ &$\to$ &$(0,\infty)$	& $ D_A = k_5/k_6A$ \\
9      	 & 0.6699		 &	$(k_{17},k_{16})$	&$\to$	&$(0,\infty)$	& $ P_C = k_{15}/k_{16}C$ \\
10     & 0.7144		 &	$(k_{9},k_{7})$	&$\to$	&$(0,\infty)$	&  $ P_B = k_9/k_{10}B$ \\
11     & 0.8544		 &	$ (k_4,k_1)$ &$\to$ &$\infty$	& $ P_A = k_3/k_4A$ \\
12     & 0.8670		 &	$(k_3,k_1)$	&$\to$	&$(0,\infty)$	& \\
13     & 0.8698		 &	$(k_{16},k_{7})$	&$\to$	&$(0,\infty)$	&  \\ 
14     & 1.6083		 &	$(k_{12},k_{8})$	&$\to$	&$\infty$	&  \\
15     & 1.6094		 &	$(k_{8},k_{7})$	&$\to$	&$\infty$	&  \\ 
16     & 2.2361		 &	$(k_{2},k_{1})$	&$\to$	&$\infty$	&  \\\hline
\multicolumn{5}{r|}{FINAL OUTPUT} & $ C_{eq} = \frac{ k_8 k_{10} k_{11} k_{13} k_{15} k_{18}}{k_7 k_9 k_{12} k_{14} k_{16} k_{17}}$
\end{tabular}
\end{table}

Steps of the second variant of the reduction of this model are given in Tab. \ref{tab:details_IFFL2}. For this variant, we took $\epsilon = 0.01$ in the fitness function and slightly different initial parameters. Final equations for the second variant of the IFFL reduction are given by:

\begin{table}[]
\centering
\caption{Incoherent feedforward loop second variant initial parameters}
\label{tab:param_IFFL2}
\begin{tabular}{c|c}
Parameter & Value \\ \hline
$ k_1$ & 3 \\
$ k_2$ & 0.5 \\
$ k_3$ & 3 \\
$ k_4$ & 1 \\
$ k_5$ & 2 \\
$ k_6$ & 3 \\
$ k_7$ & 5 \\
$ k_8$ & 1 \\
$ k_9$ & 3.5 \\
$ k_{10}$ & 1 \\
$ k_{11}$ & 5 \\
$ k_{12}$ & 1 \\
$ k_{13}$ & 2 \\
$ k_{14}$ & 1 \\
$ k_{15}$ & 3 \\
$ k_{16}$ & 2 \\
$ k_{17}$ & 1 \\
$ k_{18}$ & 3 \\
\end{tabular}
\end{table}

\begin{table}[h!]
\centering
\caption{Incoherent feedforward loop second variant}
\label{tab:details_IFFL2}
\begin{tabular}{c|c|c c c|l}
Step & $\phi_{ init}$ & Parameters        &       & Limit         & Description per group   \\ \hline
1       & 0.8788           & $ (k_{14}, k_{13})$ 	& $\to$ & $\infty$       &    \\ 
2	 & 0.8789           & $ (k_{5},k_{6})$    & $\to$ & $\infty$ &   \\ 
3	 & 0.8793           & $ (k_{9},k_{10}) $   & $\to$ & $\infty$ &    \\
4	 & 0.8802           & $ (k_{3},k_{4})$   & $\to$ & $\infty$     &      \\ 
5	 & 0.8811   	 & $ (k_{4},k_{1})$  & $\to$ & $\infty$ &     \\
6	 & 0.8814	          & $(k_{16}, k_{15})$		  & $\to$ & $\infty$    &   \\ 
7	 & 0.8822		  & $(k_{18},k_{15})$               & $\to$ &  $0,\infty$     & $P_C = D_C = 1$   \\ 
8      	 & 1.0372		 &	$(k_{11},k_{12})$ &$\to$ &$\infty$	&  \\
9      	 & 1.0375		 &	$k_{17}$	&$\to$	&$\infty$	&  \\
10     & 1.0378		 &	$(k_{2},k_{1})$	&$\to$	&$\infty$	&   \\
11     & 1.3659		 &	$ (k_6,k_8)$ &$\to$ &$0,\infty$	& $ D_A = 1$ \\
12     & 1.3736		 &	$k_{12}$	&$\to$	&$0$	& $D_B = 1$ \\
13     & 1.4156		 &	$(k_{10},k_{1})$	&$\to$	&$0,\infty$	& $P_B = 1$  \\ \hline
\multicolumn{5}{r|}{FINAL OUTPUT} & $ C_{eq} = \frac{ k_8 k_{13}}{k_7 k_{14}}$
\end{tabular}
\end{table}

\begin{align*}
 \dot{A} &=  k_1 I \, P_A - k_2 A \\
 \dot{P}_A &=  k_3 (1 - P_A) - k_4 A P_A \\
 \dot{B} &=  k_7 A  - k_8 B \\
 \dot{C} &=  k_{13} A - k_{14} B C\\
\end{align*}

The main difference between the two variants is on the equations for $B$ and $C$.  On the example of Fig. 3B in the paper, $B$ and $C$ regulate their own production rate, and $B$ titrates $C$, while on the system above, and $B$ regulates the degradation rate of $C$.

\section{Adaptive Sorting}
We perform parameter reduction on the Adaptative Sorting model without any symmetry breaking process.
Initial equations for the adaptive sorting model are
given by

\begin{align*}
 \dot{K} &=  \beta (K_T - K) - \alpha K C_0 \\
 \dot{C}_0 &=  \kappa (L- \sum_i C_i) (R - \sum_i C_i) + b C_1 - (\phi K + \tau^{-1}) C_0 \\
 \dot{C}_1 &=  \phi K C_0 - (\tau^{-1} + b) C_1.
\end{align*}

Initial parameters are given in Tab. \ref{tab:param_AS}.
Steps of the reduction of this model are given in Tab. \ref{tab:details_AS}.

\begin{table}[h]
\centering
\caption{Adaptative sorting initial parameters}
\label{tab:param_AS}
\begin{tabular}{c|c}
Parameter & Value \\ \hline
$ \phi$ & $ 3 \times 10^{-4}$ \\
$ K_T$ & $10^3$ \\
$ \alpha $ & 1 \\
$ \beta $ & 1 \\
$ \kappa $ & $ 10^{-4}$ \\
$ R$ & $ 10^4$\\
$ b$ & $ 5 \times 10^{-2}$ \\
\end{tabular}
\end{table}

\begin{table}[h]
\centering
\caption{Adaptive sorting}
\label{tab:details_AS}
\begin{tabular}{c|c|c c c|l}
Step & $I_{ init}$ & Parameters        &       & Limit         & Description per group   \\ \hline
1    & 0.8131         & $ (\alpha, \beta)$ & $\to$ & $\infty$      & $ C^* = \beta / \alpha$  \\
2	 & 0.8131		  & $ (K_T, \phi)$     & $\to$ & $(0, \infty)$ &	$ A = \phi K_T$          \\
3	 & 0.8131		  & $ (\kappa, R)$     & $\to$ & $(0, \infty)$ &	$ \kappa R \to \infty$   \\
4	 & 0.8131		  & $ R$               & $\to$ & $\infty$      &                         \\ 
5	 & 0.8645		  & $ (C^*, A)$		  & $\to$ & $(0, \infty)$ & $ \lambda = A C^*$       \\
6	 & 1		      & $ \alpha$		  & $\to$ & $\infty$      &	To undo the effect $ C_1 \propto L$ for $L \leq 2$ \\
7	 & 1		      & $ b$               & $\to$ & 0             & Uncluttering $\tau$     \\ \hline
\multicolumn{5}{r|}{FINAL OUTPUT} & $C_1 = \lambda \tau = \phi K_T \beta \tau / \alpha$
\end{tabular}
\end{table}

\section{SHP-1 model} 

\subsection{SHP-1 model first reduction}\label{sec:SHP1}
We first perform parameter reduction on the SHP-1 model with global symmetry breaking. Initial equations for the SHP-1 model are given by

\begin{align*}
 S &=  \alpha C_1 (S_T - S) - \beta S \\
 \dot{C}_0 &=  \kappa (L- \sum_i C_i) (R - \sum_i C_i) + \gamma_1 S C_1 - (\phi_1 + \tau^{-1}) C_0 \\
 \dot{C}_1 &=  \phi_1 C_0 + \gamma_2 S C_2 - (\gamma_1 S + \phi_2 + \tau^{-1}) C_1 \\
 \dot{C}_2 &=  \phi_2 C_1 + \gamma_3 S C_3 - (\gamma_2 S + \phi_3 + \tau^{-1}) C_2 \\
 \dot{C}_3 &=  \phi_3 C_2 + \gamma_4 S C_4 - (\gamma_3 S + \phi_4 + \tau^{-1}) C_3 \\
 \dot{C}_4 &=  \phi_4 C_3 + \gamma_5 S C_5 - (\gamma_4 S + \phi_5 + \tau^{-1}) C_4 \\
 \dot{C}_5 &=  \phi_5 C_4 - (\gamma_5 S + \tau^{-1}) C_5.
\end{align*}

Initial parameters for this model are given in Tab. \ref{tab:param_SHP}. Steps of the first reduction of this model are given in Tab. \ref{tab:details_SHP1}. The final system is given by the following equations when the reduction steps of Tab. \ref{tab:details_SHP1} are applied.

\begin{table}[h]
\centering
\caption{SHP-1 model initial parameters}
\label{tab:param_SHP}
\begin{tabular}{c|c}
Parameter & Value \\ \hline
$ \phi$ & $ 9 \times 10^{-2}$ \\
$ \gamma $ & $1$ \\
$ S_T $ & $7.2\times10^{-1}$ \\
$ \beta $ & $3 \times 10^{2}$ \\
$ \alpha $ & $1$ \\
$ \beta/\alpha = C^*$ & $ 3 \times 10^2$ \\
$ \kappa $ & $ 10^{-4}$ \\
$ R$ & $ 3\times 10^4$\\
\end{tabular}
\end{table}

\begin{table}[!h]
\centering
\caption{SHP-1 First reduction}
\label{tab:details_SHP1}
\begin{tabular}{c|c|c c c|l}
Step & $I_{ init}$ & Parameters             &       & Limit         & Description per group   \\ \hline
1	 & 0.7369		  & $ (\kappa, R)$         		& $\to$ & $(0,\infty)$    &      \\
2	 & 0.7369		  & $ \gamma_1$          		& $\to$ & $ \infty $       &       \\
3	 & 0.8468		  & $ (\phi_2, \phi_1)$   		& $\to$ & $(0,\infty)$   &        \\ 
4	 & 0.8583		  & $ R $    			    	& $\to$ & $\infty$	  &        \\ \hline
5	 & 0.8583		  & $ (\gamma_4,\gamma_5)$	& $\to$ & $0,\infty$   &  Kinetic sensing module      \\
6	 & 1.0000		  & $ \gamma_2 $   			& $\to$ & $ 0 $ 		  &   Uncluttering $\tau$     \\
7	 & 1.0000		  & $ \gamma_3 $   			& $\to$ & $ 0 $  	  &        \\ \hline
8	 & 1.0000		  & $ (\phi_3,S_T) $   		& $\to$ & $ \infty $	  &  Rescaling      \\
9	 & 1.0000		  & $ (\phi_1,\phi_4)$	   		& $\to$ & $ \infty $      &        \\ \hline
10	 & 1.0000		  & $ \beta $	  		  	& $\to$ & $ \infty $  	  & Adaptation module    \\
11	 & 1.0000		  & $ (\phi_4,S_T)$	   		& $\to$ & $ \infty $	   & 	    \\
12	 & 1.0000		  & $ (S_T, \alpha)$	   		& $\to$ & $ (0,\infty) $  & \\ \hline
\multicolumn{5}{r|}{FINAL OUTPUT} & $ C_5 = \frac{\phi_2 \phi_5 \beta}{\gamma_5 S_T \alpha }\tau$
\end{tabular}
\end{table}

\begin{align}
 S &=  \alpha C_1 S_T - \beta S \\
 \dot{C}_0 &=  \kappa R (L - \sum_i C_i)  + \gamma_1 S C_1 - \phi_1 C_0 \\
 \dot{C}_1 &=  \phi_1 C_0 - (\phi_2 + \gamma_1 S)C_1 \\
 \dot{C}_2 &=  \phi_2 C_1 - \phi_3 C_2 \\
 \dot{C}_3 &=  \phi_3 C_2 - \phi_4 C_3 \\
 \dot{C}_4 &=  \phi_4 C_3 + \gamma_5 S C_5 - (\phi_5 + \tau^{-1}) C_4 \\
 \dot{C}_5 &=  \phi_5 C_4 - \gamma_5 S C_5.
\end{align}

\subsection{SHP-1 model second reduction}\label{sec:SHP2}

We then perform another reduction of the same model using a different binning for the computation of the mutual information. Initial parameters and equations are identical as in the previous reduction presented in section \ref{sec:SHP1}. Steps for this reduction are given in Tab. \ref{tab:details_SHP2}. 

\begin{table}[!h]
\centering
\caption{SHP-1 Second reduction}
\label{tab:details_SHP2}
\begin{tabular}{c|c|c c c|l}
Step & $I_{ init}$ & Parameters             &       & Limit         & Description per group   \\ \hline
1	 & 0.6328		  & $ (\kappa, R)$         		& $\to$ & $(0,\infty)$    &      \\
2	 & 0.6328		  & $ R$          		& $\to$ & $ \infty $       &       \\
3	 & 0.6375		  & $ (\gamma_4, \alpha)$   		& $\to$ & $(0,\infty)$   &        \\ 
4	 & 0.6464		  & $ (\gamma_2,\gamma_1)$    			    	& $\to$ & $0,\infty$	  &        \\ \hline
5	 & 0.7264		  & $ \gamma_5$	& $\to$ & $\infty$   &  Adaptive module      \\
6	 & 1.0000		  & $ \gamma_3 $   			& $\to$ & $ 0 $ 		  &       \\
7	 & 1.0000		  & $ (\phi_1,\beta) $   			& $\to$ & $ \infty $  	  &        \\ \hline
8	 & 1.0000		  & $ \phi_4 $   		& $\to$ & $ \infty $	  &  Kinetic sensing module      \\
9	 & 1.0000		  & $ (\phi_3,S_T)$	   		& $\to$ & $ \infty $      &        \\ \hline
10	 & 1.0000		  & $ (S_T,\beta) $	  		  	& $\to$ & $ \infty $  	  &    \\
11	 & 1.0000		  & $ (\phi_5,\beta)$	   		& $\to$ & $ (0,\infty) $	   & 	    \\
12	 & 1.0000		  & $ (\beta, \phi_2)$	   		& $\to$ & $ (0,\infty) $  & \\ \hline
\multicolumn{5}{r|}{FINAL OUTPUT} & $ C_5 = \frac{\phi_2 \phi_5 \beta}{\gamma_5 S_T \alpha }\tau$
\end{tabular}
\end{table}

The final system is given by the following equations when the reduction steps given in Tab. \ref{tab:details_SHP2} are applied.

\begin{align}
 S &=  \alpha C_1 S_T - \beta S \\
 \dot{C}_0 &=  \kappa R (L - \sum_i C_i)  + \gamma_1 S C_1 - \phi_1 C_0 \label{eq:SHP2_C0} \\
 \dot{C}_1 &=  \phi_1 C_0 - (\phi_2 + \gamma_1 S) C_1 \label{eq:SHP2_C1} \\
 \dot{C}_2 &=  \phi_2 C_1 - \phi_3 C_2 \\
 \dot{C}_3 &=  \phi_3 C_2 + \gamma_4 S C_4 - \phi_4 C_3 \\
 \dot{C}_4 &=  \phi_4 C_3 + \gamma_5 S C_5 - (\phi_5 + \gamma_4 S + \tau^{-1}) C_4 \\
 \dot{C}_5 &=  \phi_5 C_4 - \gamma_5 S C_5.
\end{align}

\subsection{SHP-1 model third reduction}\label{sec:SHP3}

We perform another reduction of the same model using slightly different initial parameter values. All parameters are given in Tab. \ref{tab:param_SHP} with $S_T \to 5 S_T$. Initial set of equations is identical as in Sections \ref{sec:SHP1} and \ref{sec:SHP2}. Steps for this reduction are given in Tab. \ref{tab:details_SHP3}. 

\begin{table}[!h]
\centering
\caption{SHP-1 Third reduction}
\label{tab:details_SHP3}
\begin{tabular}{c|c|c c c|l}
Step & $I_{ init}$ & Parameters             &       & Limit         & Description per group   \\ \hline
1	 & 0.4946		  & $ (\beta, \alpha)$         		& $\to$ & $\infty$    &      \\
2	 & 0.4946		  & $ (R,\kappa)$          		& $\to$ & $(0,\infty) $       &       \\
3	 & 0.4946		  & $ (\gamma_1, \gamma_5)$   		& $\to$ & $0$   & Kinetic sensing module         \\ \hline
4	 & 1.0000		  & $ (\kappa,\phi_1)$    			    	& $\to$ & $\infty$	  &        \\
5	 & 1.0000		  & $ \phi_1$	& $\to$ & $\infty$   &       \\
6	 & 1.0000		  & $ (\phi_2,\phi_4) $   			& $\to$ & $ (0,\infty) $ 		  &       \\ \hline
7	 & 1.0000		  & $ (\phi_5,\gamma_3) $   			& $\to$ & $ \infty $  	  &   Adaptation module     \\ 
8	 & 1.0000		  & $ \gamma_2 $   		& $\to$ & $ 0 $	  &       \\
9	 & 1.0000		  & $ S_T $	   		& $\to$ & $ \infty $      &        \\ 
10	 & 1.0000		  & $ \gamma_4 $	  		  	& $\to$ & $ 0 $  	  &    \\ \hline
11	 & 1.0000		  & $ (\phi_4,\phi_3)$	   		& $\to$ & $ (0,\infty) $	   & Rescaling	    \\
12	 & 1.0000		  & $ (\alpha, \gamma_3)$	   		& $\to$ & $ (0,\infty) $  & \\ \hline
\multicolumn{5}{r|}{FINAL OUTPUT} & $ C_5 = \frac{\phi_2 \phi_3 \phi_4 \beta}{\gamma_3 S_T \alpha }\tau^2$
\end{tabular}
\end{table}

The final system is given by the following equations when the reduction steps given in Tab. \ref{tab:details_SHP3} are applied.

\begin{align*}
 S &=  \alpha C_1 S_T - \beta S \\
 \dot{C}_0 &=  \kappa (L - \sum_i C_i)(R - \sum_i C_i) - \phi_1 C_0 \\
 \dot{C}_1 &=  \phi_1 C_0 - (\phi_2 + \tau^{-1}) C_1 \\
 \dot{C}_2 &=  \phi_2 C_1 + \gamma_3 S C_3 - (\phi_3 + \tau^{-1})C_2 \\
 \dot{C}_3 &=  \phi_3 C_2 +  - \gamma_3 S C_3 \\
 \dot{C}_4 &=  \phi_4 C_3 - \phi_5 C_4 \\
 \dot{C}_5 &=  \phi_5 C_4 - \tau^{-1} C_5.
\end{align*}

\subsection{SHP-1 model reduction without feedback}

\begin{figure}
\includegraphics[width=\textwidth]{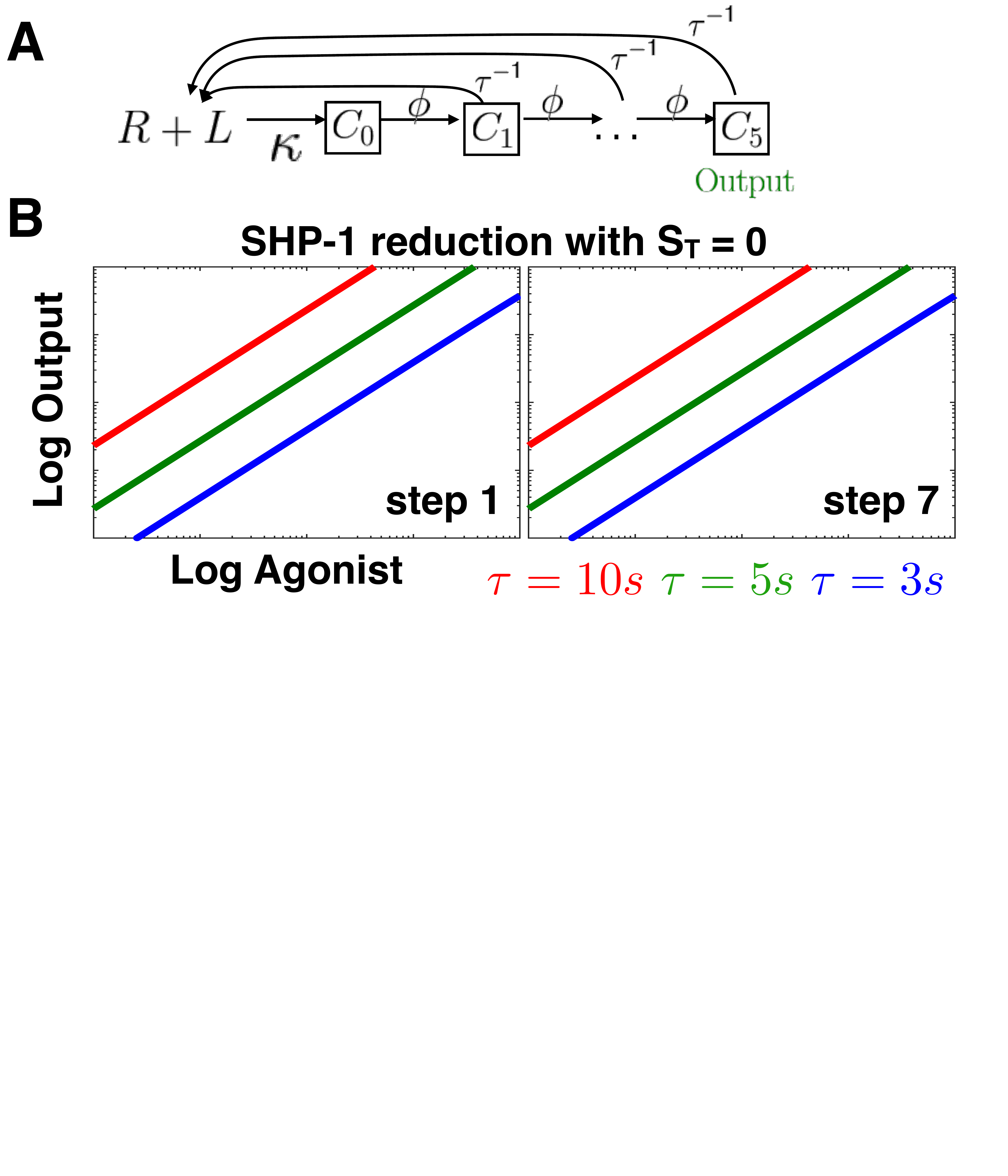}
\caption{``Negative control'' for SHP-1 model: we attempt to reduce this model with $\bar \phi$ in absence of SHP-1 (corresponding to pure kinetic proofreading). The algorithm fails to optimize behavior and fitness, indicating that it is not possible to do so for arbitrary networks.
}
\label{fig:SHP_off}
\end{figure}

We perform a reduction of the SHP-1 model with the SHP-1 mediated feedback turned off. Parameter values are given in Tab. \ref{tab:param_SHP} with $S_T = 0$. The network topology is as in Fig. S  \ref{fig:SHP_off}A and the corresponding initial set of equations is identical as in Sections \ref{sec:SHP1} and \ref{sec:SHP2}. Fig. S  \ref{fig:SHP_off}B shows that the reduction does not converge when crucial network elements (SHP-1 feedback) are missing.

\subsection{Analytical study}

The full analytical study of this model is done in \cite{Francois:2013}. Assuming all $\phi_i=\phi$ and $\gamma_i=\gamma$ are equal, we get at lowest order

\begin{equation}C_1\simeq  r_-(1-r_-)\frac{\kappa R L}{\kappa R + \nu_1}\end{equation}

with 
\begin{equation}
r_\pm=\frac{\phi+S+\nu_1\pm\sqrt{(\phi+S+\nu_1)^2-4\phi S}}{2S}
\end{equation}.

We can use the previous expression to get a closed equation for $S$ as a function of $r_-(S)$ and $C_*$.

\begin{equation}
S=S_T\frac{C_1}{C_1+C_*}=S_T\frac{ r_-(1-r_-)}{ r_-(1-r_-)+\frac{C_*(\kappa R + \nu_1)}{\kappa R L}}\label{Sanalytic}
\end{equation}

This is a 4th order polynomial equation in $S$ in terms of the parameters that can be conveniently solved numerically. Once this is done, we get the following expression for $C_N$, the final complex in the cascade as a function of $r_\pm$ to the lowest order in $r_-^{N}$.

\begin{equation}
C_N\simeq \frac{\kappa R L}{\kappa R + \nu_1}\left(1-\frac{r_-}{r_+}\right)r_-^{N}\label{CN:anal}
\end{equation}

To see why this feedback hinders perfect adaptation, it is useful to consider the limit of big $L$ and big $S_T$.  In this limit, it is shown in  \cite{Francois:2013} that the parameter $r_-$ becomes inversely proportional to the feedback variable $1/S$, thus giving at lowest order a $S^{-N}$ contribution in Eq. \ref{CN:anal}, clearly coming from the coupling of $N$ identical proofreading steps. Those equations can be approximately solved   \cite{Francois:2013} so that

\begin{equation}
C_N\simeq \left( \frac{\phi \beta}{\alpha \gamma S_T} \right)^ {N/2}(L)^{1-N/2}.
\label{intermRegim}
\end{equation}

So we see that, unless $N=2$, there is an unavoidable $L$ dependency.  The $L^{-N/2}$ dependency comes from the steady state value of the feedback variable $S \propto L^{1/2}$ appearing when we fully close this system.

\section{Lipniacki model}


In this section, we present two parameter reductions performed by $\bar{\phi}$. Initial equations for the Lipniacki model are:

\begin{align*}
 \dot{X}_2 &=  b_1 \, pMHC_{free} \, TCR_{free} + (s_2 + s_3) X_{23} - (lb \, LCK_{free} + ly_1 X_{29} + \tau^{-1}) X_2 \\
 \dot{X}_3 &=  lb \, LCK_{free} \, X_2 + ls_1 X_4 + (s_2 + s_3) X_{24} - (ly_2 + X_{37} + ly_1 X_{29} + \tau^{-1}) X_3 \\
 \dot{X}_4 &=  X_{37} X_3 - (ly_2 + ls_1 + \tau^{-1}) X_4 \\
 \dot{X}_5 &=  ly_2 X_3 + ls_1 X_6 - (tp + X_{37} + ly_1 X_{29} + \tau^{-1}) X_5 \\
 \dot{X}_6 &=  ly_2 X_4 + X_{37} X_5 - (tp + ls_1 + \tau^{-1}) X_6 \\
 \dot{X}_7 &=  tp X_5 + ls_1 X_8 - (tp + X_{37} + ly_1 X_{29} + \tau^{-1}) X_7 \\
 \dot{X}_8 &=  tp X_6 + X_{37} X_7 - (tp + ls_1 + \tau^{-1}) X_8 \\
 \dot{X}_9 &=  tp X_7 + ls_1 X_8 - (\tau^{-1}  + X_{37} + ly_1 X_{29} + \tau^{-1}) X_9 \\
 \dot{X}_{10} &=  tp X_8 + X_{37} X_9 - (ls_1+ \tau^{-1}) X_{10} \\
 \dot{X}_{22} &=  ly_1 X_{29} \, TCR_{free} + \tau^{-1} (X_{23} + X_{24}) - (s_2 + s_3) X_{22} \\
 \dot{X}_{23} &=  ly_1 X_{29} X_2 - (s_2 + s_3+ \tau^{-1}) X_{23} \\
 \dot{X}_{24} &=  ly_1 X_{29} (X_3 + X_5 + X_7 + X_9) -  (s_2 + s_3 + \tau^{-1}) X_{24} \\
 \dot{X}_{29} &=  s_1 (X_5 + X_7 + X_9) \, SHP_{free} + s_3 (X_{22} + X_{23} + X_{24}) + s_0 \, SHP_{free} \\
			 &\,\,\,\,\,  - ly_1 (X_2 + X_3 + X_5 + X_7 + X_9 + TCR_{free}) X_{29} - s_2 X_{29} \\
 \dot{X}_{31} &=  z_1 (X_9 + X_{10}) (m_1 - X_{31}) + z_0 m_1 - (z_0 + z_2) X_{31} \\
 \dot{X}_{33} &=  2 X_{31} (e_1 - X_{34}) + 2 m_2 X_{34} - (m_2 + 3 X_{31}) X_{33} \\
 \dot{X}_{34} &=  X_{31} X_{33} - 2 m_2 X_{34} \\
 \dot{X}_{36} &=  2 X_{34} (ls_2 - X_{37}) + 2 e_2 X_{37} - (e_2 +  2X_{34}) X_{36} \\
 \dot{X}_{37} &=  X_{34} X_{36} - 2 e_2 X_{37} \\
\end{align*}

\newpage

To ensure physical behavior throughout the reduction process, we manually implement the following mass conservation laws.

\begin{align*}
 pMHC_{free} &=  pMHC - \left( \sum_{i=2}^{10}X_i + X_{23} + X_{24} \right) \\
 TCR_{free} &=  TCR - \left( \sum_{i=2}^{10}X_i + X_{22} + X_{23} + X_{24} \right) \\
 LCK_{free} &=  LCK - \left( \sum_{i=3}^{10}X_i  + X_{24} \right) \\
 SHP_{free} &=  SHP - (X_{22} + X_{23} + X_{24} + X_{29}) \\
 ZAP_{free} &=  ZAP - X_{31} \\
 MEK_{free} &=  MEK - (X_{33} + X_{34}) \\
 ERK_{free} &=  ERK - (X_{36} + X_{37}) 
\end{align*}

We also perform initial rescaling of equations $ X_{31}$ to $  X_{37}$ to save $\bar{\phi}$ steps:
\begin{align*}
 X_{31} &\to  \frac{m_1 X_{31}}{ZAP} \\
 X_{33} &\to  \frac{e_1 X_{33}}{MEK} \\
 X_{34} &\to  \frac{e_1 X_{34}}{MEK} \\
 X_{36} &\to  \frac{ls_2 X_{36}}{ERK} \\
 X_{37} &\to  \frac{ls_2 X_{37}}{ERK} 
\end{align*}

Initial parameters are given in Tab. \ref{tab:param_LIP}.

\begin{table}[ht]
\centering
\caption{Lipniacki model initial parameters}
\label{tab:param_LIP}
\begin{tabular}{c|c|c}
Parameter & Value & Details\\ \hline
$ TCR$ & $ 3 \times 10^{4}$ & \\
$ LCK $ & $10^5$ & \\
$ SHP $ & $3\times10^{5}$ & \\ \hline
$ ZAP $ & $10^5$ & \\
$ MEK $ & $ 10^{5}$ & Can't be modified by $\bar{\phi}$\\ 
$ ERK $ & $ 3\times10^{5}$ & \\ \hline
$ b_1$ & $ 3\times 10^{-1}/TCR$ & Agonist peptide binding\\
$ lb$ & $  3\times 10^{-1}/LCK$ & LCK(s) binding \\
$ ly_1$ & $ 5/SHP$ & pSHP complex binding\\
$ ly_2$ & $ 3\times 10^{-1}$ & Theorine phosphorylation at complex\\
$ ls_1$ & $ 10^{-1}$ & Spontaneous serine dephosphorylation \\
$ ls_2$ & $ 5 \times 10^{-1}$ & ppERK catalyzed serine phosphorylation\\
$ tp$ & $ 5 \times 10^{-2}$ & TCR phosphorylation\\
$ s_0$ & $ 10^{-5}$ & Spontaneous SHP phosphorylation \\
$ s_1$ & $ 3 \times 10^{2}/SHP$ & SHP phosphorylation\\
$ s_2$ & $ 6 \times 10^{-4}$ & SHP dephosphorylation\\
$ s_3$ & $ 5 \times 10^{-2}$ & SHP dissociation\\
$ z_0$ & $ 2 \times 10^{-6}$ & Spontaneous ZAP phosporylation \\
$ z_1$ & $ 5/ZAP$ & ZAP phosphorylation\\
$ z_2$ & $ 2 \times 10^{-2}$ & ZAP dephosphorylation\\
$ m_1$ & $ 5\times ZAP/MEK$ & MEK phosphorylation\\
$ m_2$ & $ 2 \times 10^{-2}$ & MEK dephosphorylation\\
$ e_1$ & $ 5\times MEK/ERK$ & ERK phosphorylation\\
$ e_2$ & $ 2 \times 10^{-2}$ & ERK dephosphorylation\\
\end{tabular}
\end{table}

\subsection{Lipniacki model reduction: first variant}

For this reduction, we used mutual information as a fitness function. We discarded all values of the output below the measurable threshold $10^{-2}$, and used 40 log-uniformly distributed bins on the interval $[10^{-2},10^2]$ for the computation of the Output distribution. The Input concentrations were given by 50 log-uniformly distributed values on the interval $[1,10^4]$.

Steps of the first biochemical reduction of this model (odeLIPbasic.m in the MATLAB code) are given in Tab. \ref{tab:details_LIPbasic1}. The results of the biochemical reduction are given by

\begin{table}[h]
\centering
\caption{Lipniacki basic first variant}
\label{tab:details_LIPbasic1}
\begin{tabular}{c|c|c c c|l}
Step & $I_{ init}$ & Parameters    &       & Limit         & Description per group   \\ \hline
1	 & 0.45		  & $(m_1, m _2)$ & $\to$ &	$\infty$	  & \\
2	 & 0.47		  & $b_1$		  & $\to$ &	$\infty$	  & \\
3	 & 0.47		  & $(lb,  LCK)$	  & $\to$ &	$(0, \infty)$ &	$lb \sp{\prime} = lb  \, LCK$ \\ \hline
4	 & 0.47		  & $ls_1, z_1$	  & $\to$ &	$0, \infty$   &	Turning off the positive feedback \\
5	 & 0.50		  & $e_1, e_2$	  & $\to$ &	0	          & \\	
6	 & 0.50		  & $z_0, z_2$	  & $\to$ &	0	          & \\
7	 & 0.50		  & $m_1$		  & $\to$ &	0  	          & \\ \hline
8	 & 0.50		  & $s_3$		  & $\to$ &	0 	          & \\
9	 & 0.50		  & $(ls_2,  LCK)$ & $\to$ & $(0, \infty)$ & \\
10	 & 0.50		  & $ly_2$		  & $\to$ &	$\infty$	  & \\
11	 & 0.50		  & $( TCR,  SHP)$  & $\to$ &	$\infty$      & \\
12	 & 0.5017		  & $s_0$		  & $\to$ &	0	          & \\ \hline
13	 & 0.5017		  & $(s_2, ly_1)$ & $\to$ & $(0, \infty)$ &	Products \\
14	 & 0.5017		  & $( SHP, s_1)$  & $\to$ &	$(0, \infty)$ &	$s_1\sp{\prime} = s_1  SHP$ \\
15	 & 0.5017		  & $( LCK, ly_1)$ & $\to$ &	$(0, \infty)$ & \\
16	 & 0.5216		  & $s_1$		  & $\to$ &	$\infty$      &
\end{tabular}
\end{table}

\begin{align*}
 \dot{X}_2 &=  b_1 \, pMHC_{free} \, TCR + s_{22} X_{23} - lb X_2 \\
 \dot{X}_3 &=  lb X_2 + s_{23} X_{24} - ly_{21} X_3 \\
 \dot{X}_4 &=  X_{37} X_3 - ly_{22} X_4 \\
 \dot{X}_5 &=  ly_{21} X_3 - (tp_1 + X_{37} + ly_{14} X_{29} + \tau^{-1}) X_5 \\
 \dot{X}_6 &=  ly_{22} X_4 + X_{37} X_5 - (tp_2 + \tau^{-1}) X_6 \\
 \dot{X}_7 &=  tp_1 X_5 - (tp_3 + X_{37} + ly_{15} X_{29} + \tau^{-1}) X_7 \\
 \dot{X}_8 &=  tp_2 X_6 + X_{37} X_7 - (tp_4 + \tau^{-1}) X_8 \\
 \dot{X}_9 &=  tp_3 X_7 - (\tau^{-1}  + X_{37} + ly_{16} X_{29}) X_9 \\
 \dot{X}_{10} &=  tp_4 X_8 + X_{37} X_9 - \tau^{-1} X_{10} \\
 \dot{X}_{23} &=  ly_{12} X_{29} X_2 - (s_{22} + \tau^{-1}) X_{23} \\
 \dot{X}_{24} &=  (ly_{13} X_3 + ly_{14} X_5 + ly_{15} X_7 + ly_{16} X_9) X_{29} -  (s_{23} + \tau^{-1}) X_{24} \\
 \dot{X}_{29} &=  s_{11} X_5 + s_{12} X_7 + s_{13} X_9 - ly_{11} TCR \, X_{29}.
\end{align*}

We then perform global symmetry breaking (odeLIPadvanced in the MATLAB code). Steps of reduction are given in Tab. \ref{tab:details_LIPadvanced1}.

\begin{table}[h]
\centering
\caption{Lipniacki advanced first variant}
\label{tab:details_LIPadvanced1}
\begin{tabular}{c|c|c c c|l}
Step & $I_{ init}$ & Parameters          &       & Limit          & Description per group   \\ \hline
1	 & 0.5837		  & $(s_{22}, s_{23})$   & $\to$ & 0			 & \\
2	 & 0.5837		  & $(b_1, ly_{22})$     & $\to$ & $\infty$	     & $ly_{22} \to \infty$ makes no change \\
3	 & 0.5837		  & $(s_{21}, ly_{22})$  & $\to$ & $\infty$      &	\\ 
4	 & 0.5837		  & $ly_{22}$	  		 & $\to$ & $\infty$      &	\\ \hline
5	 & 0.5837		  & $( TCR, ly_{11})$ & $\to$ & $(0, \infty)$ &	$ly_{11}\sp{\prime} = ly_{11}  TCR$ \\
6	 & 0.5837		  & $s_{12}$	         & $\to$ &	0	         & \\
7	 & 0.6097		  & $s_{13}$	         & $\to$ &	0  	         & \\
8	 & 0.6147		  & $(tp_2, tp_3)$       & $\to$ & $(0, \infty)$ & \\
9	 & 0.6231		  & $(ly_{13}, lb)$      & $\to$ & $0, \infty$   & \\ \hline
10	 & 0.6245		  & $(tp_4, ls_2)$       & $\to$ & $(0, \infty)$ & Products \\
11	 & 0.6246		  & $(tp_3, ly_{16})$    & $\to$ & $(0, \infty)$ & \\
12	 & 0.6354		  & $(ly_{16}, ly_{15})$  & $\to$ & $(0, \infty)$ & \\
13	 & 0.6563		  & $(ly_{15}, ly_{13})$  & $\to$ & $(0, \infty)$ & \\ 
14	 & 0.6699		  & $ly_{21}$		     & $\to$ & $\infty$   & \\
15	 & 0.6749		  & $ls_2, ly_{12}$      & $\to$ & $0, \infty$ & \\
16	 & 0.7405		  & $(ly_{14}, s_{11})$	 & $\to$ & $\infty$      &
\end{tabular}
\end{table}

Global symmetry breaking results in the following system.

\begin{align*}
 \dot{X}_2 &=  b_1 \, pMHC_{free} \, TCR - lb X_2 \\
 \dot{X}_3 &=  lb X_2 - ly_{2} X_3 \\
 \dot{X}_5 &=  ly_{2} X_3 - ly_{13} X_{29} X_5 \\
 \dot{X}_7 &=  tp_1 X_5 - (tp_2 + ly_{14} X_{29}+ \tau^{-1}) X_7 \\
 \dot{X}_9 &=  tp_2 X_7 - (ly_{15} X_{29}+ \tau^{-1}) X_9 \\
 \dot{X}_{23} &=  ly_{12} X_{29} X_2 - \tau^{-1} X_{23} \\
 \dot{X}_{24} &=  ly_{13} X_{29} X_5 -  \tau^{-1} X_{24} \\
 \dot{X}_{29} &=  s_1 X_5 - ly_{11} TCR \, X_{29}
\end{align*}

Only four more steps of reduction are needed to reach perfect adaptation, namely $(ly_{13},ly_{15}) \to 0$, $(ly_{11},ly_{14}) \to \infty$, $(tp_1,ly_{14}) \to \infty$ and finally $ly_{14} \to \infty$. We apply those steps of reduction by hand and reach the final following system.

\begin{align}
 \dot{X}_2 &=  b_1 \, pMHC_{free} \, TCR - lb X_2 \label{LipV1b}\\
 \dot{X}_3 &=  lb X_2 - ly_{2} X_3 \\
 \dot{X}_5 &=  ly_{2} X_3 - tp_1 X_5 \\
 \dot{X}_7 &=  tp_1 X_5 - ly_{14} X_{29} X_7 - tp_2 X_7\\
 \dot{X}_9 &=  tp_2 X_7 - \tau^{-1} X_9 \\
\dot{X}_{24} &= ly_{14} X_7 X_{29} - \tau^{-1} X_{24} \\
\dot{X}_{29} &= s_1 X_5 - ly_{11} TCR \, X_{29} \label{LipV1e}
\end{align}

\subsection{Lipniacki model reduction: second variant}

Initial equations, parameters, mass conservation laws and equation transformations for this reduction are the same as for the previous Lipniacki reduction. For this reduction, we chose mutual information as the fitness with 40 bins log-uniformly distributed on the interval $[10^{-2},10^2]$, plus a lower bin for concentrations below $10^{-2}$ and a higher bin for concentrations above $10^2$. We chose 50 log-uniformly distributed Input concentrations on the interval  $[1,10^4]$. Because of the binning choice, the fitness, was optimized quicker, while most reduction took place in the neutral fitness landscape of maximum fitness of $1 \> bit$. The details of this biochemical reduction are given in Tab. \ref{tab:details_LIPbasic2}.

\begin{table}[h]
\centering
\caption{Lipniacki basic second variant}
\label{tab:details_LIPbasic2}
\begin{tabular}{c|c|c c c|l}
Step & $I_{ init}$ & Parameters    &       & Limit         & Description per group   \\ \hline
1	 & 0.7583		  & $(m_2, m _1)$ & $\to$ &	$\infty$  & Shutting down positive feedback\\
2	 & 0.8337		  & $(z_2,m_1)$		  & $\to$ &	$\infty$	  & \\
3	 & 0.8337		  & $ LCK$	  & $\to$ &	$\infty$ &	 \\ 
4	 & 0.8777		  & $lb$	  &  $\to$ &	$\infty$   &	\\
5	 & 0.8777		  & $(ls_1, ly_2)$	  & $\to$ &$(0,\infty)$         & \\	
6	 & 0.8777		  & $(z_0, s_0)$	  & $\to$ &	0	          & \\
7	 & 0.8777		  & $(m_1,s_1)$		  & $\to$ &	$\infty$ 	          & \\ 
8	 & 0.8777		  & $(ly_2,z_1)$		  & $\to$ &	$(0,\infty)$ 	          & \\
9	 & 0.8915		  & $e_2$ & $\to$ & $0$ & \\
10	 & 0.8915		  & $z_1$		  & $\to$ &	$0$	  & \\
11	 & 0.8915		  & $e_1$  & $\to$ &	$\infty$      & \\ \hline
12	 & 0.8915		  & $(s_1, SHP)$		  & $\to$ &	$(0,\infty)$      & Rescaling \\ 
13	 & 0.8954		  & $(b_1, SHP)$ & $\to$ & $\infty$ & \\
14	 & 0.9029		  & $(s_3, s_2)$  & $\to$ &	$(0, \infty)$ & \\
15	 & 0.9278		  & $(ls_2, tp)$ & $\to$ &	$(0, \infty)$ & \\
16	 & 0.9351		  & $(tp, TCR)$		  & $\to$ &	$(0,\infty)$      & \\
17	 & 0.9725            & $(s_2, ly_1)$  & $\to$ &	$(0, \infty)$ & \\
18	 & 1       		  & $({SHP}, ly_1)$  & $\to$ &	$(0, \infty)$ &
\end{tabular}
\end{table}

After the first reduction, the system is reduced to

\begin{align*}
 \dot{X}_2 &=  b_1 \, pMHC_{free} \, TCR_{free} + (s_{22} + s_{32}) X_{23} -  lb \, LCK \, X_2 \\
 \dot{X}_3 &=  lb \, LCK \, X_2 + ls_{11} X_4 + (s_{23} + s_{33}) X_{24} - (ly_{21} + X_{37} + ly_{11} X_{29} + \tau^{-1}) X_3 \\
 \dot{X}_4 &=  X_{37} X_3 - (ly_{22} + ls_{11} + \tau^{-1}) X_4 \\
 \dot{X}_5 &=  ly_{21} X_3 + ls_{12} X_6 - (tp_1 + X_{37} + ly_{12} X_{29} + \tau^{-1}) X_5 \\
 \dot{X}_6 &=  ly_{22} X_4 + X_{37} X_5 - (tp_2 + ls_{12} + \tau^{-1}) X_6 \\
 \dot{X}_7 &=  tp_1 X_5 + ls_{13} X_8 - (tp_3 + X_{37} + ly_{13} X_{29} + \tau^{-1}) X_7 \\
 \dot{X}_8 &=  tp_2 X_6 + X_{37} X_7 - (tp_4 + ls_{13} + \tau^{-1}) X_8 \\
 \dot{X}_9 &=  tp_3 X_7 + ls_{14} X_8 - (X_{37} + ly_{14} X_{29} + \tau^{-1}) X_9 \\
 \dot{X}_{10} &=  tp_4 X_8 + X_{37} X_9 - (ls_{14} + \tau^{-1}) X_{10} \\
 \dot{X}_{22} &=  ly_{15} X_{29} \, TCR_{free} + \tau^{-1} (X_{23} + X_{24}) - (s_{21} + s_{31}) X_{22} \\
 \dot{X}_{23} &=  ly_{16} X_{29} X_2 - (s_{22} + s_{32} + \tau^{-1}) X_{23} \\
 \dot{X}_{24} &=  X_{29} (ly_{11}X_3 + ly_{12}X_5 + ly_{13}X_7 + ly_{14}X_9) -  (s_{23} + s_{33} + \tau^{-1}) X_{24} \\
 \dot{X}_{29} &=  (s_{11} X_5 + s_{12} X_7 + s_{13} X_9) \, SHP +  (s_{31}X_{22} + s_{32}X_{23} + s_{33}X_{24}) \\
			 &\,\,\,\,\,  - (ly_{16}X_2 + ly_{11}X_3 + ly_{12}X_5 + ly_{13}X_7 + ly_{14}X_9 + ly_{15}TCR_{free}) X_{29} - s_{24} X_{29} \\
 X_{37} &= 0.05 \\
\end{align*}

We then perform global symmetry breaking on this system. Steps of the biochemical reduction of this model are given in Tab. \ref{tab:details_LIPadvanced2}

\begin{table}[h]
\centering
\caption{Lipniacki advanced second variant}
\label{tab:details_LIPadvanced2}
\begin{tabular}{c|c|c c c|l}
Step & $I_{ init}$ & Parameters    &       & Limit         & Description per group   \\ \hline
1	 & 1		  & $(m_{21}, ly _{22})$ & $\to$ &	0  & Cleaning unnecessary parameters\\
2	 & 1		  & $(m_1,s_{24})$		  & $\to$ &	$0$	  & \\
3	 & 1		  & $(ly_{11},ls_{13})$	  & $\to$ &	0 &	 \\ 
4	 & 1		  & $(s_{12},s_{31})$	  &  $\to$ & 0   &	\\
5	 & 1		  & $(ly_{13}, ls_{11})$	  & $\to$ &$0$         & \\	
6	 & 1		  & $(e_1, ls_{14})$	  & $\to$ &	0	          & \\
7	 & 1		  & $(tp_2,s_{33})$		  & $\to$ &	$0$ 	          & \\ 
8	 & 1		  & $(m_{22},ls_{12})$ & $\to$ & $0$ & \\
9	 & 1		  & $(z_2,s_{22})$		  & $\to$ &	$0$	  & \\
10	 & 1		  & $(s_{32},s_{13})$  & $\to$ &	$0$      & \\ 
11	 & 1		  & $ly_{16}$		  & $\to$ &	$0$      & \\ 
12	 & 1		  & $(s_{23},ly_{14})$ & $\to$ & $0$ & \\
13	 & 1       	  & $ls_2$  & $\to$ &	$\infty$ & \\ \hline
14	 & 1		  & $(s_{11}, tp_4)$  & $\to$ &	$\infty$ & Strengthening remaining reactions \\
15	 & 1		  & $(ly_{21}, ly_{15})$ & $\to$ &	$\infty$ & \\
16	 & 1		  & $(b_1, s_{21})$		  & $\to$ &	$\infty$      & \\ \hline
17	 & 1            & $tp_3$  & $\to$ &	$0$ & Turning off one output\\ \hline
18	 & 1       	  & $(lb, ly_{15})$  & $\to$ &	$\infty$ & Strengthening remaining reactions \\
19	 & 1		  & $({SHP}, s_{21})$  & $\to$ &	$\infty$ & \\
20	 & 1		  & $({LCK}, ly_{12} )$ & $\to$ &	$\infty$ & \\
21	 & 1		  & $(ly_{12},tp_4)$		  & $\to$ &	$\infty$      & \\
22	 & 1            & $(ly_{15}, tp_4)$  & $\to$ &	$\infty$ & \\
23	 & 1       	  & $tp_4$  & $\to$ &	$\infty$ & \\
24	 & 1       	  & $({TCR}, tp_1)$  & $\to$ &	$(0, \infty)$ & \\ \hline
\multicolumn{5}{r|}{FINAL OUTPUT} & \multicolumn{1}{l}{$X_{10} = \frac{tp_1 s_{21}  TCR}{s_{11} ly_{17}}\tau$}
\end{tabular}
\end{table}

We can remove equations for $ X_4$, $ X_6$, $X_{23}$ and $ X_{24}$ as they are dead ends in the network. $ X_{37}=0.5$ is held constant. The final expression of the output given in Tab. \ref{tab:details_LIPadvanced2} is extracted from remaining equations at steady-state; expanding the equations for the relevant cascade we get

\begin{align}
 \dot{X}_2 &=  b_1 \, pMHC_{free} \, TCR_{free} - lb \, LCK \, X_2 \label{X_2_2}\\
 \dot{X}_3 &=  lb \, LCK \, X_2 - ly_{21} X_3 \\
 \dot{X}_5 &=  ly_{21} X_3 - ly_{12} X_{29} X_5\\ 
 \dot{X}_7 &=  tp_1 X_5 - X_{37} X_7\\ 
 \dot{X}_8 &=  X_{37} X_7 - tp_4 X_8 \\
 \dot{X}_{10} &=  tp_4 X_8 - \tau^{-1} X_{10}\\ 
 \dot{X}_{22} &=  ly_{15} X_{29} \, TCR_{free} - s_{21} X_{22} \\
 \dot{X}_{29} &=  s_{11} X_5 \, SHP - ly_{15}TCR_{free} X_{29}\label{X_29_2}\\
 X_{37} &= 0.5
\end{align}
The output here is $X_{10}$. Variables $X_{22}$, $X_{29}$ and $X_5$ respectively correspond to variables $R_p$, $S$ and $C_5$ in the main text.
The structure of Eqs. \ref{X_2_2} to \ref{X_29_2} is clearly very similar to the equations of the previous reduction \ref{LipV1b} to \ref{LipV1e}, with a linear cascade for the second reduction  $X_2\rightarrow X_3\rightarrow X_5 \rightarrow  X_7 \rightarrow X_8 \rightarrow X_{10}$  and  $X_2\rightarrow X_3\rightarrow X_5 \rightarrow  X_7 \rightarrow X_9$ for the first reduction, modulated by a parallel loop via $X_{29}$ and $X_5$. As described in the main text, the structural difference comes from the mechanism of this loop, the first reduction giving an effective feedforward adaptive system, while the second reduction is an integral feedback mechanism.

\section{Antagonism}
The models we reduce have all captured the phenomenon of ligand antagonism, where the response of agonist ligands in the presence of high amounts of well chosen subthreshold ligands (i.e. with binding time lower than critical binding time $\tau_c$ triggering response) is antagonized. With our fitness, we have quantified and selected for absolute discrimination in the networks, and through the reduction, ligand antagonism has remained, but the hierarchy of antagonism has changed. In the simplest systems, antagonism is maximum for minimum $\tau$, while for more complex models there maximum antagonism is reached closer to threshold $\tau_c$ (see discussion in \cite{Francois:2016}). It turns out we can recover this property by adding two terms to the final reduced equations. 

\begin{figure}
\includegraphics[width=\textwidth]{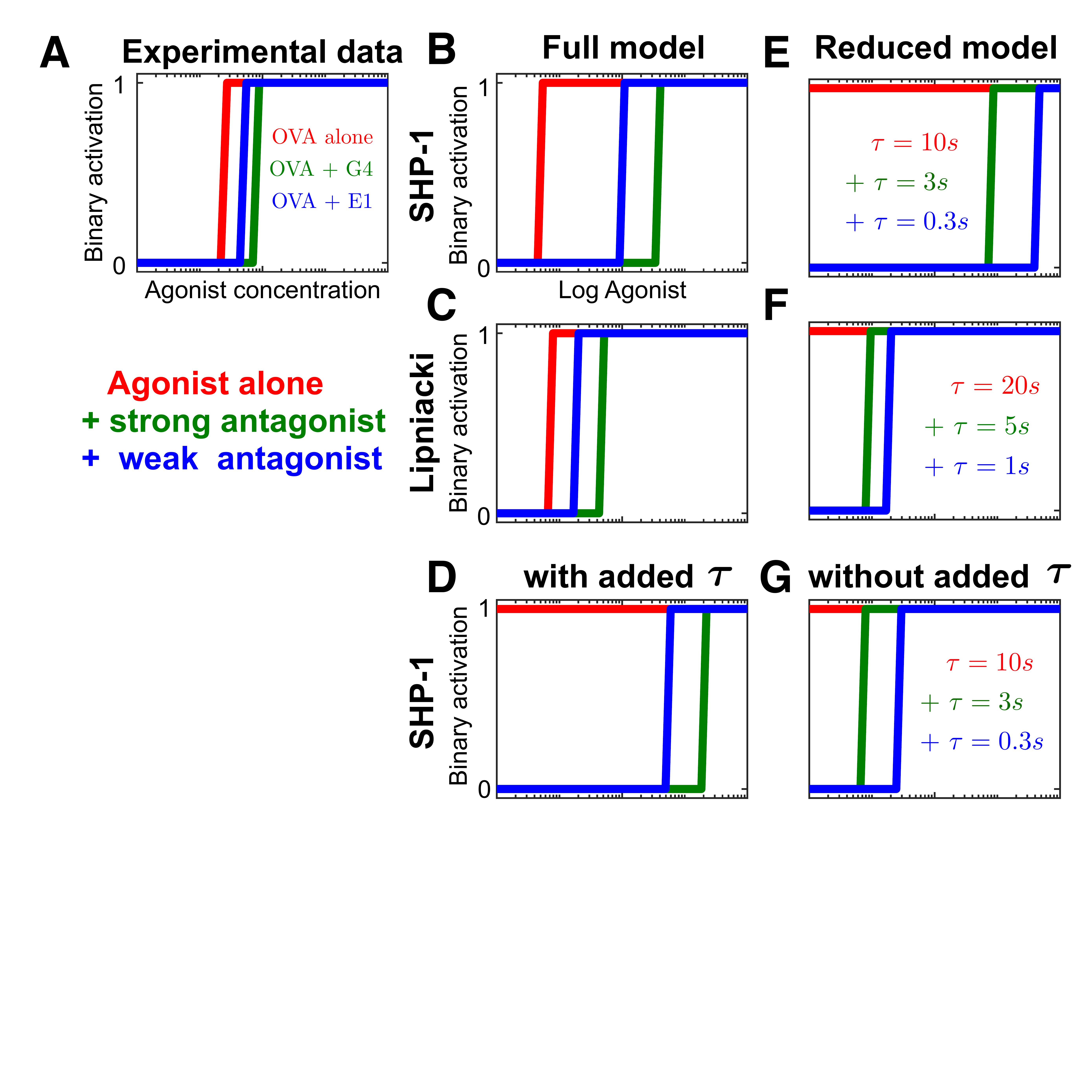}
\caption{Overview of antagonism. Red corresponds to agonists alone, green to agonists in the presence of a fixed number of strong antagonist ligands and blue to agonists with weak antagonists. The output is shown as an binary activation depending on threshold crossing. (A) Experimental data, reproduced from Fig. 5B of \cite{Francois:2013}. (B)-(C) Full SHP-1/Lipniacki model, showing typical antagonistic hierarchy with binding times as in (E),(F), which show reduced variants of the SHP-1/Lipniacki model via global symmetry breaking. (D),(G) Second variant of the reduced SHP-1 model. Upon adding back terms in $\tau$ to Eqs. \ref{eq:SHP2_C0}-\ref{eq:SHP2_C1}, we retrieve the proper hierarchy of antagonism. We added $10^4$ antagonist ligands to SHP-1 models, $10^2$ antagonist ligands to Lipniacki models and $10 \, \mu \text{mol}$ antagonist ligand concentration in the experiments.
}
\label{fig:antagonism}
\end{figure}

An overview of antagonism is presented in Fig. S\ref{fig:antagonism}. We draw the response line as a binary activation by choosing a threshold of the final output for activation (we know from our previous works \cite{Lalanne:2013, Francois:2013} that adding stochasticity to have a more probabilistic view does not fundamentally change this picture). The earliest response (in terms of ligand concentration) always comes from the agonist alone. Note how T cells in Fig. S\ref{fig:antagonism} A presented to OVA agonists + strong G4 antagonists are activated at higher agonist concentration than the weak E1 antagonists, and G4 have a binding time close to threshold than E1. This hierarchy is typical for experimentally observed antagonism: antagonism strength is large just below $\tau_c$, the critical binding time above which a response is elicited. 

Similarly, in the full models for SHP-1 and Lipniacki (Fig. S\ref{fig:antagonism} B - C), we have the same hierarchy. However, for the same binding times in reduced SHP-1 (Fig. S\ref{fig:antagonism} E) and reduced Lipniacki (Fig. S\ref{fig:antagonism} F), we have an inverted hierarchy, where ligands further below are more antagonizing, so closer to the naive models discussed in \cite{Francois:2016}. 

It turns out that the position of the adaptive module $m$ in the kinetic proofreading cascade of $N$ complexes, defined as the complex on which the variable $S$ implements the negative "tug-of-war" term described in the main text,  determines the antagonism strength, like in Fig. 4 of ref. \cite{Francois:2016}. We can rescue the correct hierarchy of antagonism by adding kinetic terms $\tau^{-1}$ to the equations. We illustrate this on the second variant of SHP-1 reduction. The antagonism hierarchy is initially absent from the reduced model (Fig. S\ref{fig:antagonism} G). When we add $\tau^{-1}$ terms  to Eqs. \ref{eq:SHP2_C0} and \ref{eq:SHP2_C1}, it is retrieved, Fig. \ref{fig:antagonism}, because  $m = 4$ is large enough. When $m$ is too low ($m = 2$, Figs. S\ref{fig:antagonism} E - F), antagonism behavior peaks for $\tau \ll \tau_c$ and we can not recover the hierarchy observed experimentally.

